# Pseudophase-Change Effects in Turbulent Channel Flow under Transcritical Temperature Conditions

**Kukjin Kim[1], Jean-Pierre Hickey[2]† and Carlo Scalo[1]**

[1]School of Mechanical Engineering, Purdue University, 585 Purdue Mall, West Lafayette, IN 47907-2088, USA

[2]Department of Mechanical and Mechatronics Engineering, University of Waterloo, 200 University Avenue West, Waterloo, ON N2L 3G1, Canada

We have performed direct numerical simulations (DNS) of compressible turbulent channel flow at supercritical pressure with top and bottom isothermal walls kept respectively at a supercritical ($T_{top} > T_{pb}$) and subcritical temperature ($T_{bot} < T_{pb}$), where $T_{pb}$ is the pseudoboiling temperature. The DNS are conducted using a high-order discretization of the fully compressible Navier–Stokes equations in conservative form closed with the Peng-Robinsion (PR) state equation. Bulk density is adjusted to obtain a bulk pressure of approximately $p_b = 1.1 p_{cr}$ where $p_{cr}$ is the critical pressure of the working fluid. Top-to-bottom temperature differences investigated are $\Delta T$ = 5 K, 10 K, and 20 K, where $T_{top/bot} = T_{pb} \pm \Delta T/2$; buoyancy effects are neglected. Varying $\Delta T$ modifies the average location of pseudophase change from $y_{pb}/h = -0.23$ ($\Delta T$ = 5 K) to 0.89 ($\Delta T$ = 20 K), where $h$ is the channel half-height and $y = 0$ the centerline position. Real-fluid effects cause visible deviations from classical scaling laws in the mean velocity profile. Enstrophy generation due stretching and tilting decreases with $\Delta T$. The proximity to the pseudotransitioning layer inhibits the intensity of the velocity fluctuations, while enhancing the density and temperature fluctuations. Conditional probability analysis reveals that the sheet of fluid undergoing pseudophase change is characterized by a dramatic reduction in the kurtosis of density fluctuations and becomes thinner as $\Delta T$ is increased. Instantaneous visualizations show dense fluid ejections from the pseudoliquid viscous sublayer, some reaching the channel core, causing positive values of density skewness in the respective buffer-layer region (vice versa for the top wall).

## 1. Introduction

The operating pressure of propulsion and energy systems, such as gas turbines, liquid rocket engines, or supercritical water-cooled reactors, is continuously increasing to improve performances. As a result, the working fluid often reaches pressures and temperatures exceeding its critical values, $p > p_{cr}$ and $T > T_{cr}$ respectively, hence achieving a supercritical state. While promoting high heat-transfer rates and thermo-dynamic efficiencies, and suppressing detrimental interfacial effects commonly found in low-pressure boiling or cavitation processes (Zhong *et al.* 2009; Zhang *et al.* 2011; Wen & Gu 2011), the heightened coupling between pressure, temperature, and density in the supercritical regime also accentuates unwanted fluid dynamic instabilities such as thermoacoustic oscillations in combustion chambers (Casiano *et al.* 2010) or in fuel heat exchangers (Thurston 1964; Palumbo 2009; Wang *et al.* 2015); the latter often lead to catastrophic hardware failure if uncontrolled. These so-called real-fluid effects are intensified in near-critical conditions, $p \sim p_{cr}$ and $T \sim T_{cr}$, which will be examined by the present manuscript in the context of turbulent heat-and-mass transfer in a canonical compressible turbulent channel flow setting.

† Email address for correspondence: jean-pierre.hickey@uwaterloo.ca

The lay understanding is that supercritical fluids share properties of both gases and liquids, in a seemingly homogeneous yet ambiguous state of matter. In reality, there is an identifiable transition between pseudoliquid (or liquid-like) and pseudogaseous (or gaseous-like) conditions, especially in the vicinity of the critical point, defined by the pseudoboiling line (PBL), also termed the Fisher–Widom line (Fisher & Widom 1969). The PBL is an extension of the subcritical gas-liquid coexistence curve above the critical point (Banuti 2015) and is hereafter defined as the locus of temperature and pressure values $(T_{pb} > T_{cr}, p_{pb} > p_{cr})$ at which the thermal expansion coefficient of the fluid, $\alpha_p = -\left(\partial \rho / \partial T\right)_p / \rho$, is maximum. A pseudo phase transition, or simply pseudotransition, occurs, for example, when temperature changes from $T < T_{pb}$ to $T > T_{pb}$ (or vice versa), for given pressure conditions $p = p_{pb}$, hence crossing the PBL in the $p-T$ phase diagram. The goal of the present work is to investigate the dynamics of turbulent heat and mass transfer when the instantaneous temperature and density fields fluctuate about such pseudoboiling conditions, also referred to here as transcritical temperature conditions.

Unlike a subcritical phase change, there is no well-defined latent heat since supercritical pseudotransition takes place progressively over a finite temperature range bracketing pseudoboiling (PB) conditions. While in the liquid-like ($T << T_{pb}$) or gas-like ($T >> T_{pb}$) supercritical states, molecules are homogeneously distributed in space with a well-defined mean free path, during pseudotransition heterogeneously distributed microscopic clusters of tightly packed molecules are formed (Tucker 1999). This results in abrupt changes in compressibility and density, and a rapid, albeit continuous, increase in the heat capacity, with gas-like behaviour retained between denser molecular clusters. This heterogeneous microscopic distribution results in optical dispersion effects allowing the experimental identification pseudotransition (Gorelli *et al.* 2006; Simeoni *et al.* 2010).

Due to the steep variations of macroscopic thermodynamic properties near the PBL, accurate simulations of flows in transcritical temperature conditions are numerically challenging. The lack of a well-defined interface does not permit the use of Lagrangian interface tracking methods. An Eulerian approach based on the fully compressible conservative Navier–Stokes equations typically results significant numerical instabilities, due to lack of appropriate spatial resolution. These issues are analogous to the ones found in Eulerian-Eulerian multiphase simulations with compressible solvers (Abgrall & Saurel 2003). To bypass such stability constraints, Terashima *et al.* (2011), Terashima & Koshi (2012, 2013), and Kawai (2016) used a nonconservative pressure-based formulation coupled with the use of artificial viscosity, successfully suppressing nonphysical numerical oscillations, at the expenses of energy conservation. Alternative approaches have used a double-flux formulation (Ma *et al.* 2017), inspired by the interfacial flow community, where where the flux at one face is computed twice, each time assuming a specific heat ratio taken alternatively from the left or right side of the flux face. Other works, such as Peeters *et al.* (2016), use a low-Mach-number formulation neglecting compressibility effects such as acoustic wave propagation with significant gains in computational time and stability. In the present study, a fully compressible and fully conservative approach is adopted, where numerical stability issues are contained via systematic grid refinement in the canonical setting of low Reynolds number channel flow turbulence. To ensure numerical stability on coarse grids, the conserved variables are explicitly filtered at every time step; The intensity of the filter is then progressively reduced as grid convergence is achieved.

Transcritical temperature conditions have been found to enhance heat transfer fluctuations and alter turbulence production rates in wall bounded flows (Yoo 2013). Such deviations from ideal gas behaviour are not to be confused with real-gas effects, which refers to molecularly disassociated gases occurring in hypersonic flows. Real-fluid effects in a flat-plate turbulent boundary layer (TBL) over a heated wall were studied by Kawai (2016); he found that Morkovin's hypothesis is not applicable in pseudophase changing conditions due to the presence of significant density fluctuations yielding nonclassical effects in the mass flux, turbulent diffusion, and pressure dilatation distributions. Patel *et al.* (2015) numerically and theoretically investigated the near-wall scaling laws in a turbulent channel flow with large thermophysical property variations. They confirmed that the turbulent flow statistics exhibit quasi-similarity based on a semi-local Reynolds number, $Re_\tau^* \equiv \sqrt{(\bar{\rho}/\rho_w)}/(\bar{\mu}/\mu_w)\, Re_\tau$, where the overbar refers to the Reynolds averaging and the subscript $w$ to the averaged wall quantity. Their investigation was, however, limited to a density ratio of $\bar{\rho}/\rho_w = 0.4$ to $1.0$. Nemati *et al.* (2015) performed direct numerical simulations (DNS) of a heated turbulent pipe flow at supercritical pressure, where thermal expansion due to a constant wall heat flux in the presence of low buoyancy effects was found to attenuate turbulent kinetic energy (TKE); turbulence enhancement was observed for high buoyancy cases. Pizzarelli *et al.* (2009) studied turbulent rectangular channel flow at supercritical pressure with high wall heat flux, finding that real-fluid effects attenuate heat transfer significantly at the channel corners. Compressible channel flow simulations at supercritical pressures and transcritical temperatures by Sengupta *et al.* (2017) show that the cold wall region has higher density and temperature fluctuations as well as higher coherence than the hot near-wall region. Also, the liquid-like flow region is characterized by decreased streamwise and increased spanwise anisotropy and vice versa in the region of gas-like behavior.

In the present paper, we analyze data from a DNS of compressible channel flow turbulence maintained in pseudophase-changing conditions by a wall-to-wall temperature differential imposed via isothermal conditions. The dataset presented here has been considerably expanded with respect to previous publications by the authors (Kim *et al.* 2017*a*,*b*) and analyzed in more depth. In the following, we first describe the governing equations, the fluid model, and the computational setup (section 2). The mean and fluctuating hydro- and thermodynamic quantities are then presented together with probability density distribution functions (sections 3 and 4). Finally, instantaneous turbulent structures are investigated and compared with the correlation statistics to infer their role in the heat-and-mass-transfer dynamics focusing on the near-wall region (section 5).

## 2. Problem formulation

### 2.1. Governing equations

The governing equations of mass, momentum, and total energy for a fully compressible flow are solved in conservative form, which reads

$$\frac{\partial \rho}{\partial t} + \frac{\partial \rho u_j}{\partial x_j} = 0 \qquad (2.1a)$$

$$\frac{\partial \rho u_i}{\partial t} + \frac{\partial \rho u_i u_j}{\partial x_j} = -\frac{\partial p}{\partial x_j} + \frac{\partial \tau_{ij}}{\partial x_j} \qquad (2.1b)$$

$$\frac{\partial \rho E}{\partial t} + \frac{\partial}{\partial x_j}\left[u_j(\rho E + p)\right] = \frac{\partial}{\partial x_j}(u_i \tau_{ij} - q_j) \qquad (2.1c)$$

where $x_1$, $x_2$, and $x_3$ (equivalently, $x$, $y$, and $z$) are the streamwise, wall-normal, and spanwise coordinates, respectively, and $u_i$ is the velocity component in the $i$-th direction, $t$ time, $\rho$ density, $p$ pressure, and $E$ the total energy per unit mass. Unless otherwise stated, all symbols refer to dimensional quantities.

The viscous and conductive heat fluxes in (2.1b) and (2.1c) are, respectively

$$\tau_{ij} = 2\mu\left[S_{ij} - \frac{1}{3}\frac{\partial u_k}{\partial x_k}\delta_{ij}\right] \qquad (2.2a)$$

$$q_j = -\lambda \frac{\partial T}{\partial x_j} = -\frac{c_p \mu}{Pr}\frac{\partial T}{\partial x_j} \qquad (2.2b)$$

where $\mu$ is the dynamic viscosity, $S_{ij}$ the strain rate tensor given by $S_{ij} = (\partial u_j/\partial x_i + \partial u_i/\partial x_j)/2$, $\lambda$ the thermal conductivity, $c_p$ the heat capacity at constant pressure, $Pr$ the Prandtl number, and $T$ the temperature.

### 2.2. Real fluid model

The Peng-Robinson (PR) equation of state (EoS) (Peng & Robinson 1976) is used to model the working fluid of choice for this study, R-134a (1,1,1,2-tetrafluoroethane, $CH_2FCF_3$), which is characterized by experimentally attainable critical pressures and temperatures of $p_{cr} = 40.59$ bar and $T_{cr} = 374.26$ K, respectively. Departure functions guaranteeing full thermodynamic consistency with the chosen EoS have been derived following Sengers *et al.* (2000). Transport properties such as viscosity and thermal conductivity are estimated via Chung's method (Chung *et al.* 1988), which predicts experimental values within 5% error (Poling *et al.* 2001). The choice of an accurate and simple equation of state such as the PR EoS provides a consistent thermodynamic model, computationally less expensive than interpolating tabulated values. Detailed derivations and comparisons against the NIST database (Lemmon *et al.* 2016) are included in Appendix A.

### 2.3. Computational setup

The proposed numerical simulations have been carried out with *Hybrid*, a fully compressible high-order (in space and time) Navier-Stokes solver originally written by Prof. Johan Larsson. This code has been used in several canonical numerical investigations such as shock-vortex interaction, compressible homogeneous isotropic turbulence (Larsson *et al.* 2007) and shock/turbulence interaction (Larsson & Lele 2009; Larsson *et al.* 2013). The code solves single-component fluid, which is a suitable modeling approach for a supercritical flow since for supercritical pressures, $p > p_{cr}$, surface tension becomes negligible and numerical techniques typical of multi-phase simulations, such as interface tracking or reconstruction, are not required. New features that have been added to the code include parallel HDF5 (The HDF Group 1998) input/output capabilities and a generic equation of state.

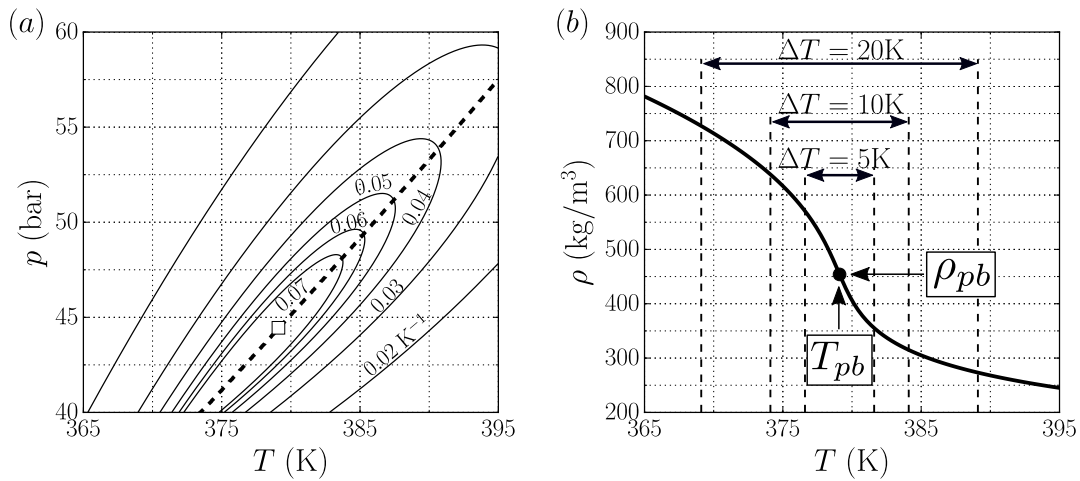

FIGURE 1. Phase diagram for R-134a showing the critical point ($p_{cr} = 40.59$ bar, $T_{cr} = 374.26$ K) (□), the pseudoboiling line (- - -), and the isolines of isobaric thermal expansion coefficient $\alpha_p$ (——) (*a*); density versus temperature for $p = 1.1 p_{cr}$ with the pseudoboiling point (●) and top-to-bottom temperature differences, $\Delta T$, bracketing the value of $T_{pb} = 379.1$ K (*b*).

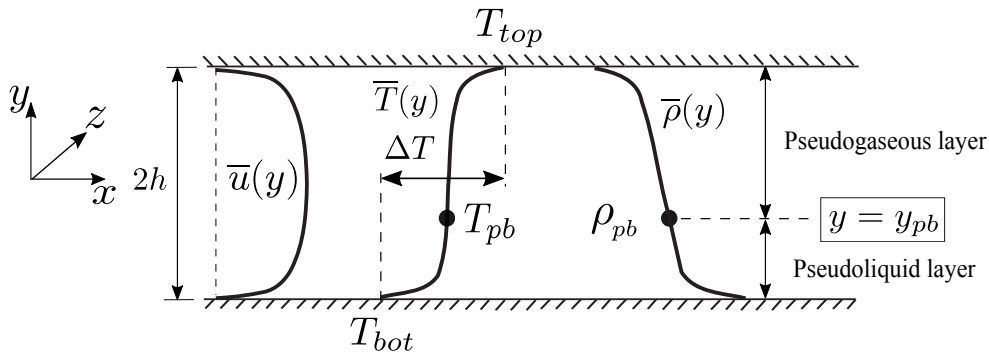

FIGURE 2. Computational setup for supercritical compressible channel flow simulations in transcritical temperature conditions. Simulation parameters are given in tables 1 and 2.

| Fluid | $p_b$ | $\rho_{pb}$ (kg/m³) | $T_{pb}$ (K) | $\Delta T$ (K) | $T_{bot}$ (K) | $T_{top}$ (K) | $\rho_b$ (kg/m³) | $U_b$ (m/s) | Box size (mm³) |
|---|---|---|---|---|---|---|---|---|---|
| R-134a ($CH_2FCF_3$) | $1.1 p_{cr}$ | 453.5 | 379.1 | 5 | 376.6 | 381.6 | 450 | 36 | 12×2×4 |
|  |  |  |  | 10 | 374.1 | 384.1 | 474 |  |  |
|  |  |  |  | 20 | 369.1 | 389.1 | 520 |  |  |

TABLE 1. Simulation parameters achieving transcritical temperature conditions for R-134a. Bulk parameters are indicated with a subscript '*b*', while pseudoboiling or pseudo(phase)transitioning with '*pb*'.

The computational setup is a compressible turbulent channel flow (figure 2) kept at a nominal bulk pressure of $p_b \simeq 1.1 p_{cr}$, corresponding to a pseudoboiling temperature of $T_{pb} = 379.1$ K defined based on the maximum thermal expansion coefficient, $\alpha_p =$ $-(\partial \rho/\partial T)_p/\rho$ (figure 1a). The assigned isothermal top and bottom wall boundary conditions bracket the pseudoboiling temperature, $T_{top/bot} = T_{pb} \pm \Delta T/2$, maintaining transcritical temperature conditions (figure 1b).

Top-to-bottom wall temperature differences investigated are $\Delta T = T_{top} - T_{bot} = 5$ K, 10 K, and 20 K, with bulk density set to $\rho_b = 450$ kg/m$^3$, 474 kg/m$^3$, and 520 kg/m$^3$, respectively, determined via trial and error to obtain the desired bulk pressure for all cases (see tables 1 and 2). Periodic boundary conditions are applied in the streamwise and spanwise directions and the grid is stretched in the wall-normal direction with a hyperbolic tangent law. To guarantee feasibility of the simulations on the finest grid and highest temperature difference considered, where the time step is acoustically limited to $\Delta t = 1.4 \times 10^{-8}$, the bulk velocity has been set for all cases to the relatively high value (for typical heat-transfer applications) of $U_b = 36$ m/s corresponding to a Mach number in the low-subsonic range of $M_b = 0.26$.

A reference ideal gas (IG) simulation is carried out with the following nondimensional parameters: $\rho_{b,*}^{(IG)} = 1.0$, $p_{b,*}^{(IG)} = 0.71$, $T_{bot,*}^{(IG)} = 0.8$, $T_{top,*}^{(IG)} = 1.2$, $\Delta T_*^{(IG)} = 0.4$, $U_{b,*}^{(IG)} = M_b = 0.26$, $Pr = 0.7$, $Re_\tau = 367$, and box size of $8 \times 2 \times 4$. The subscript "$*$" indicates nondimensional quantities scaled with the bulk density and speed of sound based on the centerline temperature. Results from the IG reference case are hereafter only presented in dimensional form, scaled based on the flow parameters of the $\Delta T = 20$ K transcritical case, which read: $\rho_{b,20K} = 520$ kg/m$^3$, $T_{pb} = 379.1$ K and $U_b = 36$ m/s, the last two being common to all cases. The rescaling relations for the IG data therefore read: $\rho_b^{(IG)} = \rho_{b,*}^{(IG)} \times \rho_{b,20K}$, $T^{(IG)} = T_{pb} + (T_*^{(IG)} - 1) \times (20\text{K}/\Delta T_*^{(IG)})$, $U_b^{(IG)} = U_{b,*}^{(IG)} \times (U_b/M_b)$.

To ensure the proper spatial resolution of all relevant thermo- and hydrodynamic scales, a systematic grid refinement study has been carried out (see Appendix B and table 2); this is especially important in simulations of supercritical flows in near-critical or pseudophase transitioning conditions (see Introduction). The relevant metric of the degree of spectral broadening for channel flow turbulence is the friction Reynolds number,

$$Re_\tau = \frac{h u_\tau}{\nu} \quad (2.3)$$

based on the friction velocity, $u_\tau$, the channel half-height, $h$, and the kinematic viscosity $\nu$ of the fluid. It can be viewed as the channel half-height normalized by the viscous length scale, $\nu/u_\tau = \nu/(\partial u/\partial x_2)_{x_2=0}$, hence $Re_\tau = h^+$. Therefore, $Re_\tau$ is the ratio of an integral length scale, $\sim h$, to a viscous scale evaluated at the wall. Typical practice in Direct Numerical Simulations (DNS) is to adopt relatively low values of friction Reynolds number to enable full resolution of the relevant scales. For the present simulations this is achieved by augmenting dynamic viscosity and thermal conductivity by the same multiplicative factor (figure 3) resulting in $Re_\tau$ in the range of 342–394 (table 2). This choice leaves the Prandtl number unaltered, and reproduces the correct trend of transport properties in the transcritical regime (see Appendix A). The IG simulations have been carried out at: $Re_\tau = 462$, $\Delta x^+ = 14.44$, $\Delta y^+ = 0.48$–9.10, $\Delta z^+ = 9.62$ for the bottom wall and $Re_\tau = 271$, $\Delta x^+ = 8.47$, $\Delta y^+ = 0.28$–5.36, $\Delta z^+ = 5.64$ for the top wall.

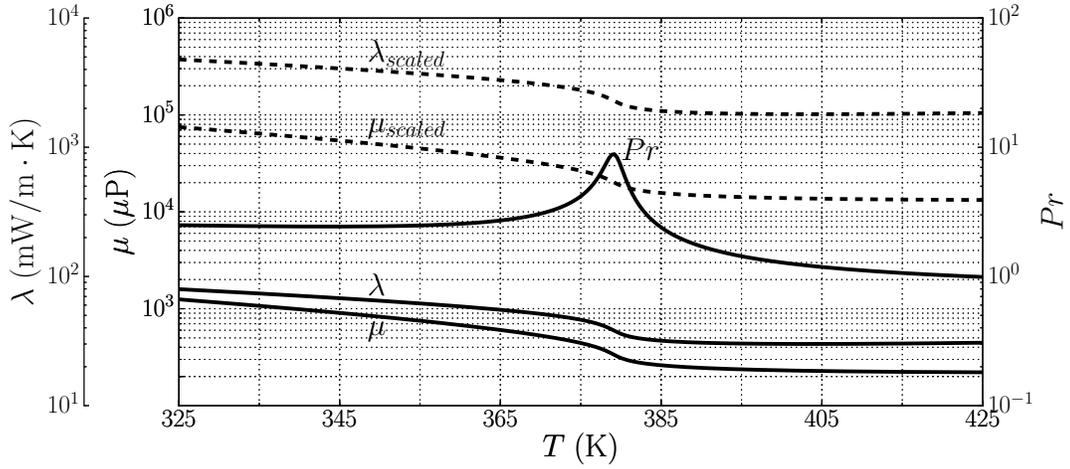

FIGURE 3. Dynamic viscosity, $\mu$, thermal conductivity, $\lambda$ and Prandtl number $Pr$ for R-134a taken from the Chung's model (———) (see Appendix A); Scaled dynamic viscosity and conductivity (- - -), augmented by a factor of 60, used in the computations, yielding the same Prandtl number.

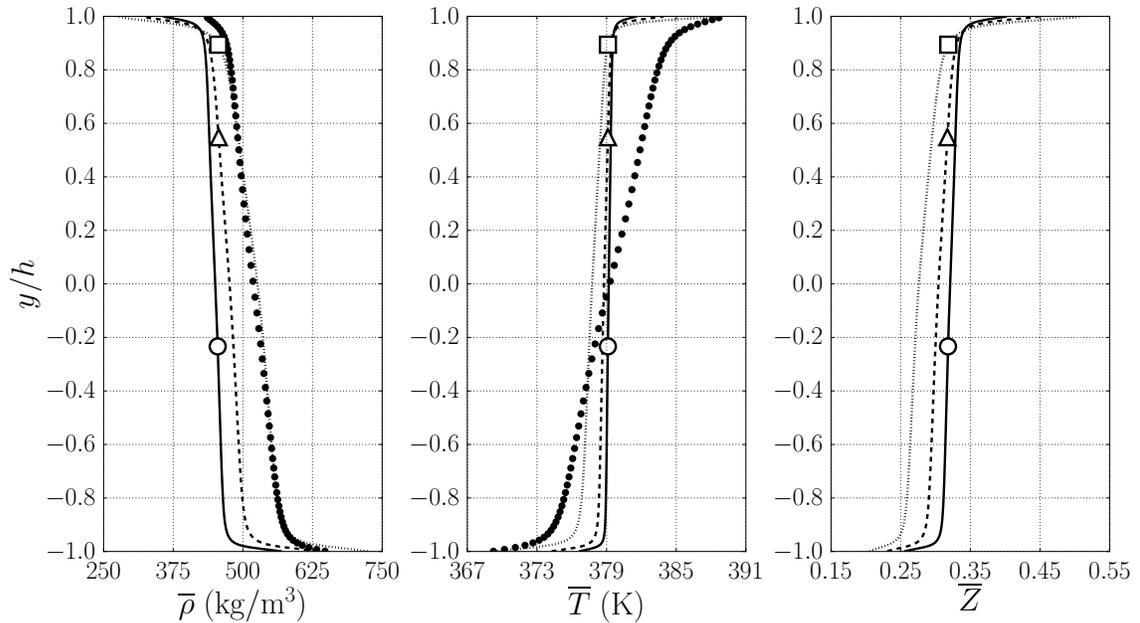

FIGURE 4. Reynolds-averaged density, temperature, and compressibility factor for $p_b = 1.1 p_{cr}$ and $\Delta T = 5$ K (———), 10 K (- - -), and 20 K ($\cdots$); rescaled ideal gas data ($\bullet$) (in section 2.3). Average location of pseudotransition for $\Delta T=$ 5 K ($\circ$), 10 K ($\triangle$), and 20 K ($\square$).

## 3. First and Second Order Statistics

In this section a statistical analysis limited to first and second-order moments of turbulent fluctuations in the transcritical channel flow setup of figure 2 is carried out in comparison with the IG simulations.

| $N_x \times N_y \times N_z$ | | 64×96×64 | 128×128×96 | 192×128×128 | 384×256×256 | 512×256×256 |
|---|---|---|---|---|---|---|
| $\Delta T = 5K, \rho_b = 450$ kg/m³ | $p_b$ | 44.64 bar | 44.65 bar | 44.67 bar | 44.66 bar | 44.67 bar |
| Bot | $Re_\tau$ | 360 | 340 | 345 | 370 | 372 |
| | $\Delta x^+$ | 67.50 | 31.88 | 21.56 | 11.56 | 8.72 |
| | $\Delta y^+$ | 0.41–16.75 | 0.40–11.03 | 0.40–11.15 | 0.39–5.09 | 0.38–5.06 |
| | $\Delta z^+$ | 22.50 | 14.17 | 10.78 | 5.78 | 5.81 |
| Top | $Re_\tau$ | 375 | 355 | 360 | 390 | 394 |
| | $\Delta x^+$ | 70.31 | 33.28 | 22.50 | 12.19 | 9.23 |
| | $\Delta y^+$ | 0.43–17.43 | 0.41–11.50 | 0.42–11.68 | 0.41–5.36 | 0.40–5.34 |
| | $\Delta z^+$ | 23.44 | 14.79 | 11.25 | 6.09 | 6.16 |
| $\Delta T = 10K, \rho_b = 474$ kg/m³ | $p_b$ | 44.58 bar | 44.65 bar | 44.65 bar | 44.67 bar | 44.69 bar |
| Bot | $Re_\tau$ | 345 | 325 | 335 | 365 | 364 |
| | $\Delta x^+$ | 64.69 | 30.47 | 20.94 | 11.41 | 8.53 |
| | $\Delta y^+$ | 0.40–16.13 | 0.38–10.65 | 0.39–10.87 | 0.38–4.96 | 0.37–4.93 |
| | $\Delta z^+$ | 21.56 | 13.54 | 10.47 | 5.70 | 5.69 |
| Top | $Re_\tau$ | 365 | 345 | 355 | 385 | 387 |
| | $\Delta x^+$ | 68.44 | 32.34 | 22.19 | 12.03 | 9.07 |
| | $\Delta y^+$ | 0.42–16.98 | 0.40–11.24 | 0.41–11.48 | 0.40–5.28 | 0.40–5.25 |
| | $\Delta z^+$ | 22.81 | 14.38 | 11.09 | 6.02 | 6.05 |
| $\Delta T = 20K, \rho_b = 520$ kg/m³ | $p_b$ | 44.37 bar | 44.43 bar | 44.42 bar | 44.55 bar | 44.67 bar |
| Bot | $Re_\tau$ | 320 | 310 | 315 | 345 | 342 |
| | $\Delta x^+$ | 60.00 | 29.06 | 19.69 | 10.78 | 8.02 |
| | $\Delta y^+$ | 0.37–15.06 | 0.36–10.03 | 0.37–10.26 | 0.36–4.72 | 0.35–4.68 |
| | $\Delta z^+$ | 20.00 | 12.92 | 9.84 | 5.39 | 5.34 |
| Top | $Re_\tau$ | 340 | 330 | 335 | 375 | 377 |
| | $\Delta x^+$ | 63.75 | 30.94 | 20.94 | 11.72 | 8.84 |
| | $\Delta y^+$ | 0.39–15.89 | 0.39–10.75 | 0.39–10.91 | 0.39–5.10 | 0.39–5.10 |
| | $\Delta z^+$ | 21.25 | 13.75 | 10.47 | 5.86 | 5.89 |

TABLE 2. Friction Reynolds number and grid resolution in wall units $(u_\tau/\nu)^{-1}$ for the bottom and top portion of the channel evaluated with respective wall quantities. See also table 1.

### 3.1. *Mean flow quantities*

Figure 4 shows Reynolds-averaged profiles of density, temperature, and compressibility factor,

$$Z = \frac{p}{\rho \, R_{gas} \, T} \qquad (3.1)$$

where $R_{gas} = 81.49$ J/kg K is the gas constant for R-134a. The top-to-bottom density difference (table 3) of $\Delta \rho = 447.5$ kg/m³ achieved under transcritical conditions for $\Delta T = 20$ K is more than twice the IG value of $\Delta \rho_{IG} = 213.5$ kg/m³ obtained for the same $\Delta T$. Departure from the ideal gas behavior is, in fact, present in the entire channel, with $\overline{Z}_{max} \simeq 0.51$ achieved at the top wall in the pseudogaseous layer for $\Delta T = 20$ K. As $\Delta T$ is increased, the average location of pseudotransition $y_{pb}$, where real-fluid effects are expected to be the most accentuated, moves from a near-centerline location to the upper wall. In all cases, the transition from a seemingly fully thermally mixed region in channel core (i.e. $\overline{T}(y)$ is relatively uniform and close to the pseudoboiling value to a conductive sublayer region at the walls is more defined than in the reference IG simulation. Such steep mean flow gradients sustain significant density and enthalpy fluctuations, up to $\rho_{rms,max} = 44.1$ kg/m$^3$ and $h_{rms,max} = 8.9$ kJ/kg, respectively (as discussed later in figures 10 and 11) for the $\Delta T = 20$ K case. The very high heat capacity of the fluid undergoing pseudo phase transition, on the other hand, limits the temperature fluctuations to $T_{rms,max} < 2$ K.

| $\Delta T$ (K) | $\overline{\rho}_{top}$ (kg/m$^3$) | $\overline{\rho}_{bot}$ (kg/m$^3$) | $\Delta\rho$ (kg/m$^3$) | $\Delta\rho_{IG}$ (kg/m$^3$) | $\overline{Z}_{top}$ | $\overline{Z}_{bot}$ | $y_{pb}$ |
|---|---|---|---|---|---|---|---|
| 5  | 358.3 | 567.7 | 209.4 | –     | 0.40 | 0.26 | $-0.23h$ |
| 10 | 318.6 | 635.0 | 316.4 | –     | 0.45 | 0.23 | $+0.55h$ |
| 20 | 276.1 | 723.6 | 447.5 | 213.5 | 0.51 | 0.21 | $+0.89h$ |

TABLE 3. Top and bottom-wall values of mean density and compressibility factor and average location of pseudophase transition $y_{pb}$ for various temperature conditions. With the exception of $\Delta T$, all values reported are a result of the calculations. $\Delta\rho_{IG}$ is obtained by rescaling the output of the (dimensionless) reference ideal gas simulation to match the flow settings of the $\Delta T = 20$ K transcritical case (in section 2.3).

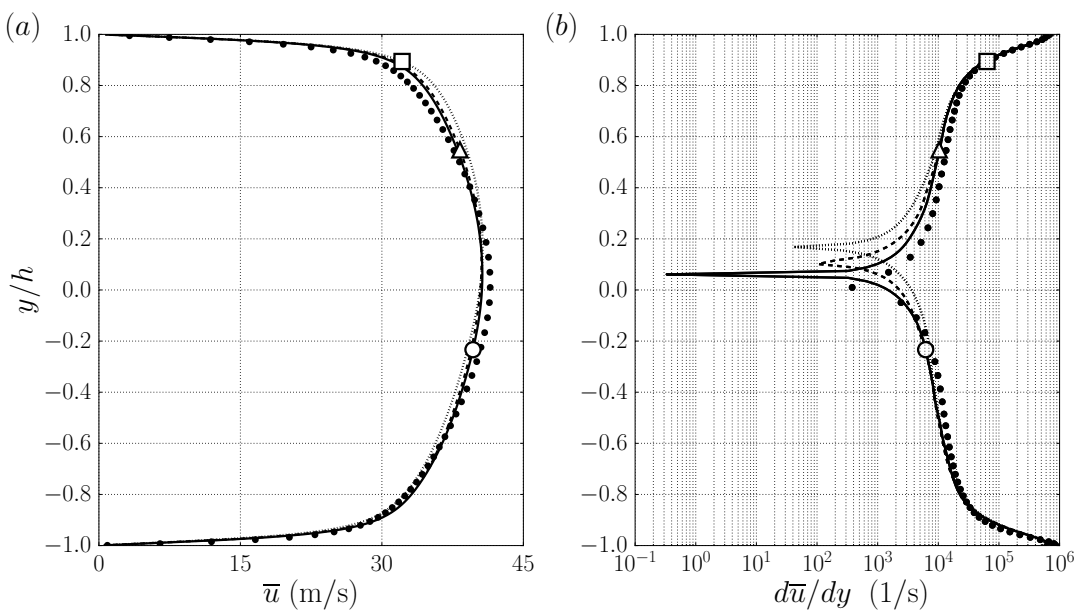

FIGURE 5. Reynolds-averaged streamwise velocity component (*a*) and shear magnitude (*b*) for $p_b = 1.1 p_{cr}$ and $\Delta T = 5$ K (——), 10 K (- - -), and 20 K (···) with reference ideal gas data (•). Average location of pseudotransition for $\Delta T = 5$ K (○), 10 K (△), and 20 K (□).

The mean turbulent streamwise velocity profile (figure 5*a*) becomes more asymmetric (with a slight acceleration of the pseudogaseous layer) for increasing $\Delta T$, with an upwards shift in the maximum velocity location, $y/h = 0.06$ for $\Delta T = 5$ K, 0.11 for $\Delta T = 10$ K, and 0.17 for $\Delta T = 20$ K, following the same trend of the pseudotransition, or boiling, location $y_{pb}$ (figure 5*b*). While top-down asymmetries in the temperature gradient are confined to the sublayer regions, the mean density gradient profile is more visibly affected by the location of pseudotransition. A logarithmic increment of the centerline of the temperature gradient is observed as $\Delta T$ is also increased logarithmically (i.e. $d(\Delta T)/\Delta T$ = const), suggesting a linear relation between the overall top-to-bottom equilibrium heat flux and $\Delta T$. The latter is a surprising result given the degree of thermodynamic and hydrodynamic nonlinearity of the problem. These results also suggest that transcritical heat flux rates are amenable to straightforward dimensionless scaling in similar canonical setups. This analysis, however, is out of the scope of the current study. While the velocity gradient increase (decrease) in the pseudogaseous (pseudoliquid) region as $\Delta T$ is increased is not as significant as the corresponding variations in density and temperature gradients, real-fluid effects are very apparent when attempting to scale the mean velocity profiles with commonly used scaling laws.

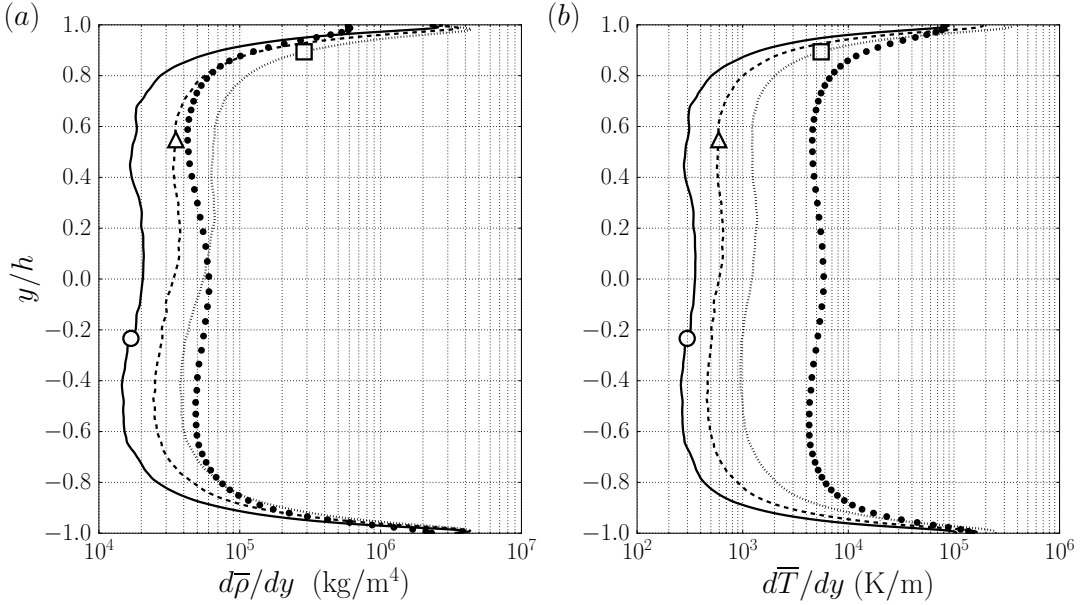

FIGURE 6. Wall-normal gradient of Reynolds-averaged mean density (*a*) and temperature (*b*) for $p_b = 1.1 p_{cr}$ and $\Delta T = 5$ K (——), 10 K (- - -), and 20 K ($\cdots$) with reference ideal gas data (●). Average location of pseudotransition for $\Delta T =$ 5 K (○), 10 K (△), and 20 K (□).

For all $\Delta T$ values, the mean streamwise velocity profiles are scaled following the recently proposed approach by Trettel & Larsson (2016), which accounts for the wall heat transfer effects, the van Driest transform (van Driest 1951, 1956), and the semi-local scaling (Huang *et al.* 1995) (figure 7). The expressions of the transformations are reported here for convenience.

The van Driest transformation (van Driest 1951, 1956) is given by

$$\overline{u}_{VD}^+ = \int_0^{\overline{u}^+} \left( \frac{\overline{\rho}(y)}{\overline{\rho}_w} \right)^{\frac{1}{2}} d\overline{u}^+ \qquad (3.2)$$

where the conventional set of scaling parameters reads

$$y^+ = \frac{y}{\delta_v} = \frac{y}{\mu_w/\overline{\rho}_w u_\tau}, \quad \overline{u}^+ = \overline{u}/\overline{u}_\tau, \quad \overline{u}_\tau = \sqrt{\overline{\tau}_w/\overline{\rho}_w} \qquad (3.3)$$

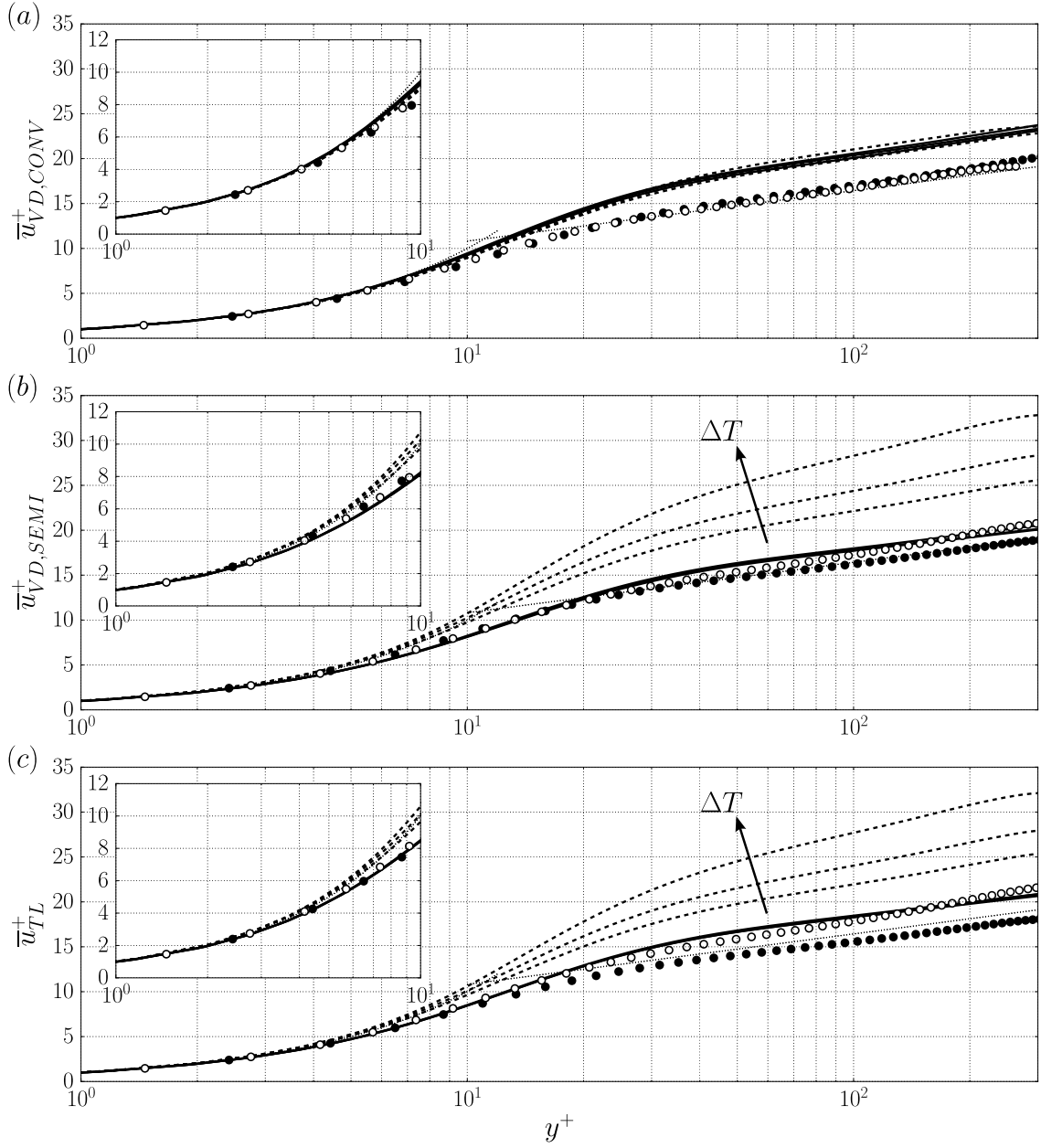

FIGURE 7. Mean streamwise velocity versus wall-normal coordinate in wall units scaled based on the conventional van Driest transformation, $\overline{u}^+_{VD,CONV}$ (3.3) (*a*), semi-local scaling $\overline{u}^+_{VD,SEMI}$ by Huang *et al.* (1995) (3.4) (*b*), and the transformation by Trettel & Larsson (2016), $\overline{u}^+_{TL}$ (*c*), for $p_b = 1.1 p_{cr}$ and $\Delta T = 5$ K, 10 K, and 20 K with reference ideal gas data (circles); Bottom wall (——, ●) and top wall (- - -, ○). The law of the wall ($\overline{u}^+ = y^+$) and the log-law ($\overline{u}^+ = \frac{1}{\kappa} \ln y^+ + C^+$ where $\kappa = 0.41$ and $C^+ = 5.2$) ($\cdots$) are shown for reference.

whereas, for the semi-local scaling by Huang *et al.* (1995) they read

$$y^+ = \frac{y}{\delta^*_v} = \frac{y}{\overline{\mu}(y)/\overline{\rho}(y) u^*_\tau}, \quad \overline{u}^+ = \overline{u}/\overline{u}^*_\tau, \quad \overline{u}^*_\tau = \sqrt{\overline{\tau}_w/\overline{\rho}(y)}. \tag{3.4}$$

Finally, the scaling by Trettel & Larsson (2016) reads

$$\overline{u}^+_{TL} = \int_0^{\overline{u}^+} \left(\frac{\overline{\rho}(y)}{\overline{\rho}_w}\right)^{\frac{1}{2}} \left[1 + \frac{1}{2}\frac{1}{\overline{\rho}(y)}\frac{d\overline{\rho}(y)}{dy}y - \frac{1}{\overline{\mu}(y)}\frac{d\overline{\mu}(y)}{dy}y\right] d\overline{u}^+ \tag{3.5}$$

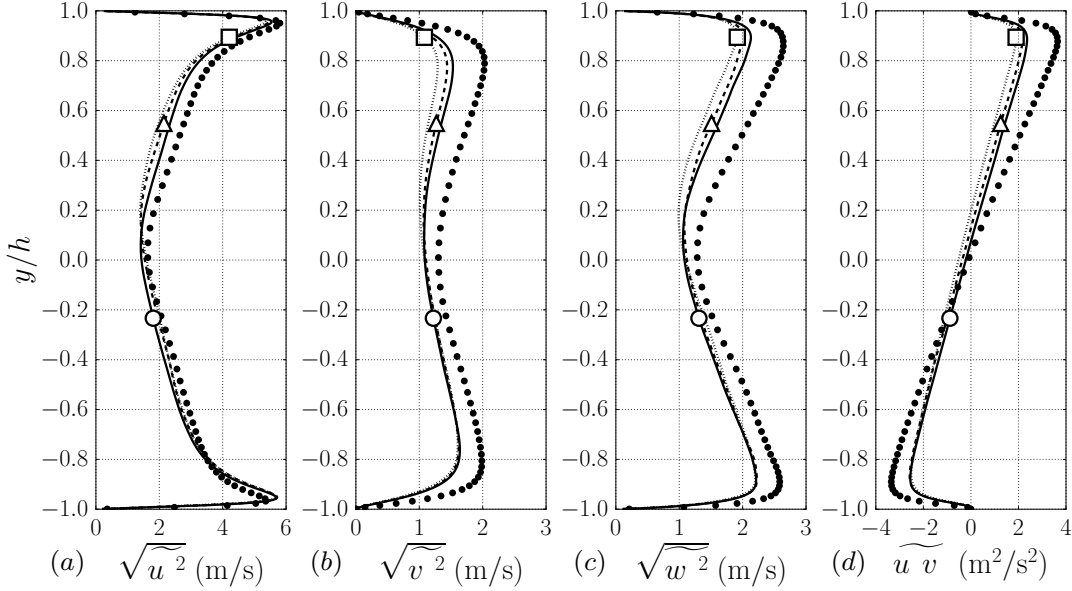

FIGURE 8. Root-mean-square of streamwise ($a$), wall-normal ($b$) and spanwise ($c$) Favre fluctuating velocity component and Reynolds shear stress ($d$) for $p_b = 1.1 p_{cr}$ and $\Delta T = 5$ K (——), 10 K (- - -), and 20 K ($\cdots$) and reference ideal gas data ($\bullet$). Average location of pseudotransition for $\Delta T$= 5 K ($\bigcirc$), 10 K ($\triangle$), and 20 K ($\square$).

The conventional van Driest transformation (figure 7$a$) collapses bottom wall and top wall velocity profiles on each other and over the range of $\Delta T$ investigated; the reference log-law intercept value is, however, only matched by the ideal gas data. On the other hand, the semi-local scaling proposed by Huang et al. (1995) is able to (approximately) collapse ideal gas and real-fluid data only for the bottom (or cooled) wall, while the dependency on $\Delta T$ of the scaled top (or heated) wall profiles is not absorbed by either Huang et al. (1995)'s or Trettel & Larsson (2016)'s transformation for this dataset. This is attributed to the strong density variations associated with pseudoboiling conditions, which migrate closer to the top wall as $\Delta T$ is increased.

The conventional van Driest velocity scaling overall leads to a better collapse over the different $\Delta T$'s, both in the viscous sublayer region (see insets in figure 7) and in the log-law region, compared to Huang's semi-local scaling, and Trettel & Larsson (2016)'s transformation. More work is required to devise correct scaling laws in the presence of very steep density and fluid properties variations in order to inform wall models and closures for the Reynolds-Averaged Navier-Stokes (RANS) equations.

### 3.2. Turbulent fluctuation intensities

Other real-fluid effects associated with transcritical thermal conditions are observable in the variance of the hydrodynamic turbulent fluctuations, as shown in figure 8. As $\Delta T$ increases, the asymmetries with respect to the channel centerline grow, with peak fluctuation intensity values at the top wall (pseudogaseous region, towards which the pseudotransition location migrates) are attenuated with respect to the corresponding values in the pseudoliquid flow (see table 4). This suggests the occurrence of damping of hydrodynamic turbulence due to the proximity to the region of pseudophase change. Such attenuation is noted in all Reynolds stress terms but is strongest in the wall-normal velocity component, directly involved in the turbulent heat and mass transport transport working against the steep mean temperature and density gradient.

| $\Delta T$ | $\Delta(u_{rms,peak})$ | $\Delta(v_{rms,peak})$ | $\Delta(w_{rms,peak})$ | $\Delta(\rho_{rms,peak})$ | $\Delta(T_{rms,peak})$ | $\Delta(p_{rms,peak})$ |
|---|---|---|---|---|---|---|
| 5 K  | −2.60 % | −6.69 %  | −3.69 %  | −31.71 % | −7.27 %  | −18.83 % |
| 10 K | −4.65 % | −11.31 % | −10.36 % | 3.90 %   | 13.37 %  | −35.44 % |
| 20 K | −5.77 % | −19.74 % | −12.72 % | 31.84 %  | 27.47 %  | −25.30 % |

TABLE 4. Top-to-bottom difference in root-mean-square peak values of streamwise, wall-normal and spanwise velocity components in percentage of the bottom peak rms value.

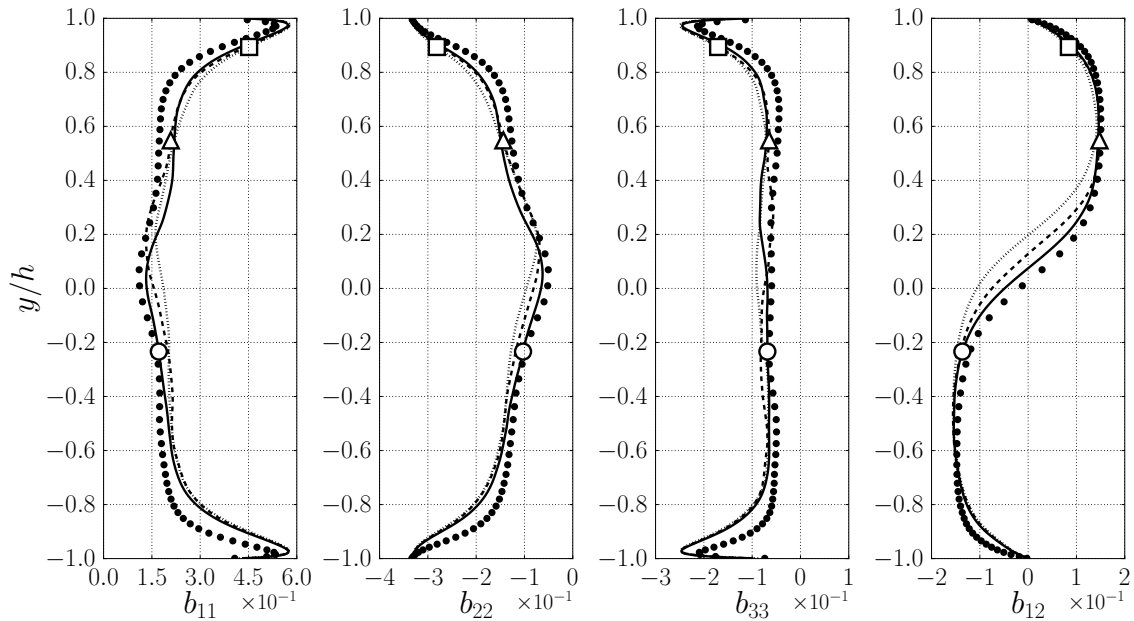

FIGURE 9. Normalized anisotropy for $p_b = 1.1 p_{cr}$ and $\Delta T = 5$ K (——), 10 K (- - -), and 20 K (· · ·) and the ideal gas data (•). Average location of pseudotransition for $\Delta T=$ 5 K (○), 10 K (△), and 20 K (□).

On the contrary, the proximity to the pseudophase change location $y_{\mathrm{pb}}$ (locally) and the increasing top-to-bottom temperature difference $\Delta T$ (globally) enhances the intensity of all thermodynamic fluctuations (figure 10 and 11*a*). In spite of the damping in the wall-normal velocity fluctuations, the wall-normal turbulent enthalpy flux is enhanced (figure 11*b*) by the increasing $\Delta T$, as expected by the statistical steadiness of the flow, implying equilibrium conditions for the turbulent heat transfer.

For any given $\Delta T$, the rms peak of density, temperature and enthalpy closer to the location of pseudophase transition $y_{\mathrm{pb}}$ has a higher value than the other one farther away. Such asymmetry is quantified in table 4. As $y_{\mathrm{pb}}$ moves upwards for increasing $\Delta T$, it approaches the peak of the shear Reynolds stress and enthalpy flux (figure 11*b*), only significantly increasing the latter roughly proportionally to $\Delta T$.

### 3.3. *Vorticity transport*

In the following, we analyze the mean vorticity and enstrophy transport equation to understand the influence of transcritical temperature conditions, responsible for large mean and instantaneous density gradients, on the turbulent velocity fluctuations.

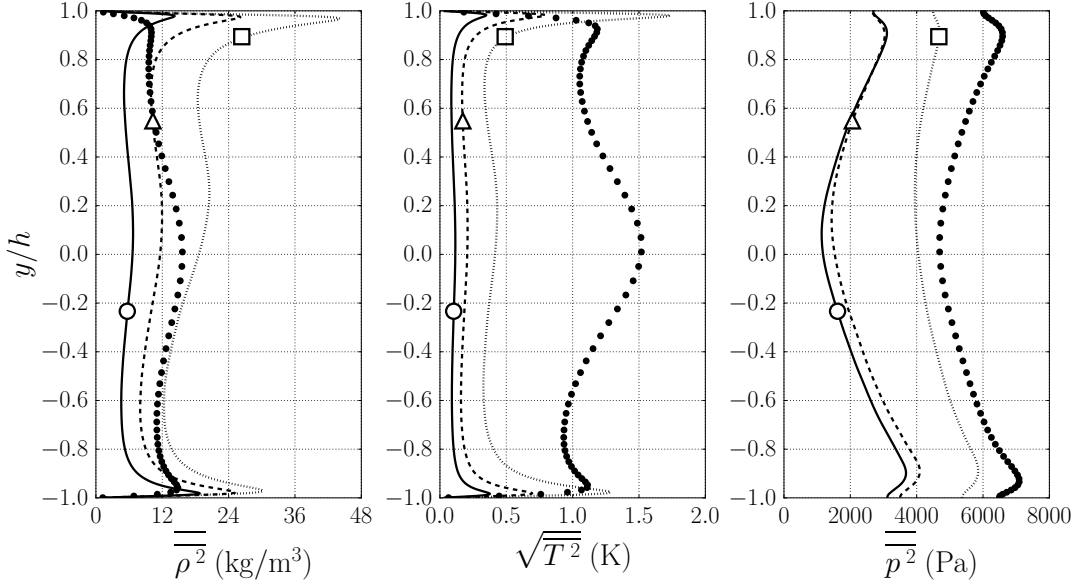

FIGURE 10. Root-mean-square of Reynolds fluctuations for density, temperature, and pressure $p_b = 1.1 p_{cr}$ and $\Delta T = 5$ K (——), 10 K (- - -), and 20 K ($\cdots$) and the ideal gas data ($\bullet$). Average location of pseudotransition for $\Delta T = 5$ K ($\bigcirc$), 10 K ($\triangle$), and 20 K ($\square$).

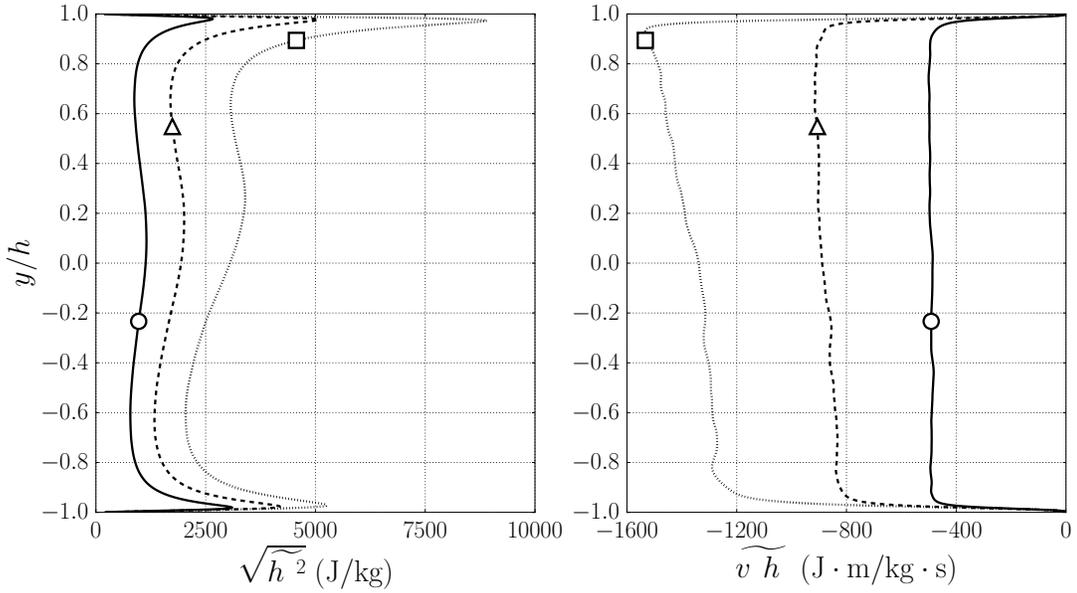

FIGURE 11. Root-mean-square of Favre fluctuations of enthalpy ($a$) and wall-normal turbulent enthalpy flux ($b$) for $p_b = 1.1 p_{cr}$ and $\Delta T = 5$ K (——), 10 K (- - -), and 20 K ($\cdots$). Average location of pseudotransition for $\Delta T = 5$ K ($\bigcirc$), 10 K ($\triangle$), and 20 K ($\square$).

The transport equation for instantaneous (or total) vorticity reads, in vector form:

$$\frac{\partial \vec{\omega}}{\partial t} + (\vec{u} \cdot \nabla) \vec{\omega} = \underbrace{(\vec{\omega} \cdot \nabla) \vec{u}}_{(I)} - \underbrace{\vec{\omega}(\nabla \cdot \vec{u})}_{(II)} + \underbrace{\frac{1}{\rho^2} \nabla \rho \times \nabla p}_{(III)} + \underbrace{\nabla \times \left( \frac{\nabla \cdot \tau}{\rho} \right)}_{(IV)} \qquad (3.6)$$

The four terms on the right-hand side of (3.6) account for ($I$) vorticity stretching and tilting by the flow velocity gradients; ($II$) vorticity stretching by the dilatational field;

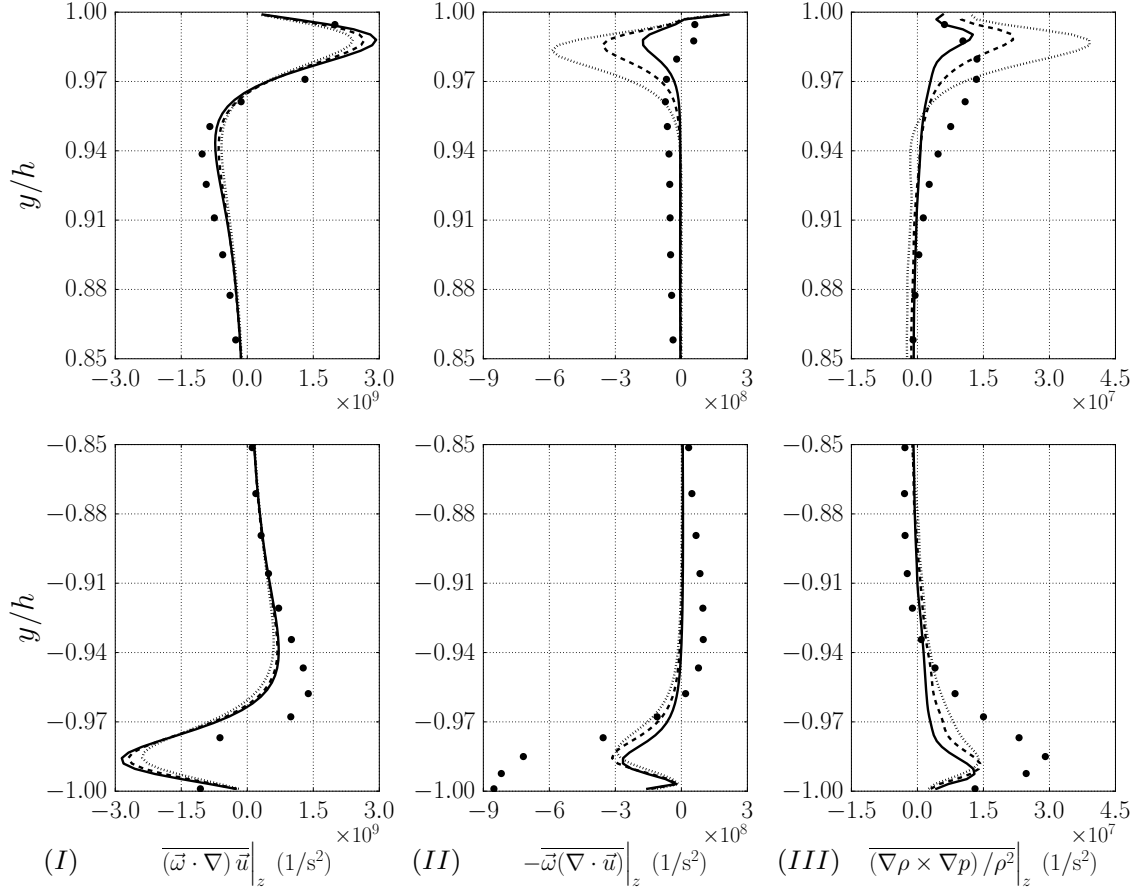

FIGURE 12. Mean spanwise vorticity budgets: effects of vorticity stretching/tilting by the flow velocity gradients $(I)$, vorticity stretching by the flow compressibility $(II)$, and the baroclinity $(III)$ for $p_b = 1.1 p_{cr}$ and $\Delta T = 5$ K (——), 10 K (- - -), and 20 K ($\cdots$). Average location of pseudotransition for $\Delta T=$ 5 K ($\bigcirc$), 10 K ($\triangle$), and 20 K ($\square$).

$(III)$ baroclinic effects; $(IV)$ vorticity diffusion by viscous effects. It is expected that transcritical temperature conditions emphasize dilatational $(II)$ and baroclinic $(III)$ effects; however, contrary to intuition, this is not the case, as discussed below.

The analysis starts with the mean vorticity transport, which is obtained by applying the Reynolds-averaging operator, isolating only the spanwise component of vorticity $\Omega_z$ and assuming frozen transport (i.e. linearized convection) yielding

$$\cancel{\frac{d\Omega_z}{dt}} + \cancel{U\frac{d\Omega_z}{dx}} = \overline{(\vec{\omega}\cdot\nabla)\vec{u}}\Big|_z - \overline{\vec{\omega}(\nabla\cdot\vec{u})}\Big|_z + \overline{\frac{1}{\rho^2}\nabla\rho\times\nabla p}\Big|_z + \overline{\nabla\times\left(\frac{\nabla\cdot\tau}{\rho}\right)}\Big|_z \quad (3.7)$$

where $U$ is the mean streamwise velocity component. Profiles of the inviscid terms of (3.7) are extracted and plotted in figure 12. As the near-wall thermal gradients are increased with increasing $\Delta T$, the absolute value of vorticity production is enhanced, as expected. Yet, in relative terms, the contribution of baroclinic torque to the mean vorticity, and hence the mean velocity profile, is one to two orders of magnitude smaller than the mean vortex stretching terms $(I)$ and $(II)$. Despite, the very strong thermodynamic effects in the flow, baroclinicity remains of limited importance for the mean flow.

The transport equation for the enstrophy evaluated based on the fluctuating vorticity field is hereafter analyzed in the continued quest to quantify the baroclinic effects triggered by the transcritical temperature conditions. A simplified form of such transport equation based on instantaneous planar averages (instead of Reynolds or Favre averages, as derived in Appendix C) is analyzed in the following. The one-time, planar average operator $\langle \cdot \rangle_{xz}$ and its associated fluctuations, indicated with the superscript ($"$) yield the following decomposition

$$\vec{\omega} = \langle \vec{\omega} \rangle_{xz} + \vec{\omega}" \qquad (3.8)$$

where to good approximation $\langle \vec{\omega} \rangle_{xz} \simeq \Omega_z$ and $\langle \vec{u} \rangle_{xz} \simeq (U, 0, 0)$. By assuming that advection of vorticity fluctuations is mainly sustained by the mean velocity:

$$\frac{\overline{D}}{\overline{D}t}\vec{\omega}" = [(\vec{\omega} \cdot \nabla) \vec{u}]" - [\vec{\omega}(\nabla \cdot \vec{u})]" + \left[\frac{1}{\rho^2}\nabla \rho \times \nabla p\right]" + \left[\nabla \times \left(\frac{\nabla \cdot \tau}{\rho}\right)\right]" \qquad (3.9)$$

where $\overline{D}/\overline{D}t = \partial/\partial t + U\partial/\partial x$. By multiplying both sides by the vorticity fluctuations $\vec{\omega}"$, the approximate transport equation for enstrophy reads

$$\frac{\overline{D}}{\overline{D}t}\zeta = \underbrace{\overline{\vec{\omega}" \cdot [(\vec{\omega} \cdot \nabla) \vec{u}]"}}_{(I)} - \underbrace{\overline{\vec{\omega}" \cdot [\vec{\omega}(\nabla \cdot \vec{u})]"}}_{(II)} + \underbrace{\overline{\vec{\omega}" \cdot \left[\frac{1}{\rho^2}\nabla \rho \times \nabla p\right]"}}_{(III)} + \underbrace{\overline{\vec{\omega}" \cdot \left[\nabla \times \left(\frac{\nabla \cdot \tau}{\rho}\right)\right]"}}_{(IV)}$$

$$(3.10)$$

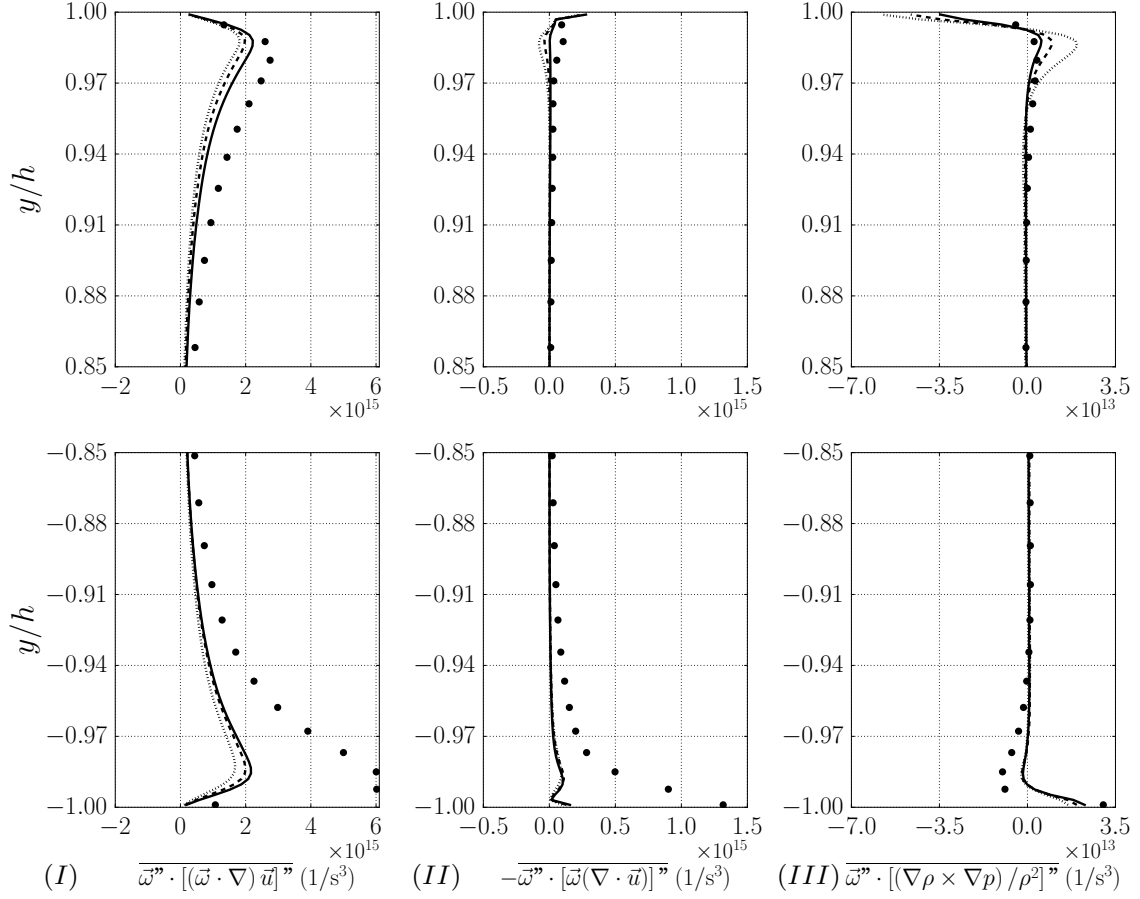

FIGURE 13. Enstrophy budgets: effects of vorticity stretching/tilting by the flow velocity gradients $(I)$, vorticity stretching by the flow compressibility $(II)$, and the baroclinity $(III)$ for $p_b = 1.1 p_{cr}$ and $\Delta T = 5$ K (——), 10 K (- - -), and 20 K ($\cdots$). Average location of pseudotransition for $\Delta T=$ 5 K ($\circ$), 10 K ($\triangle$), and 20 K ($\square$).

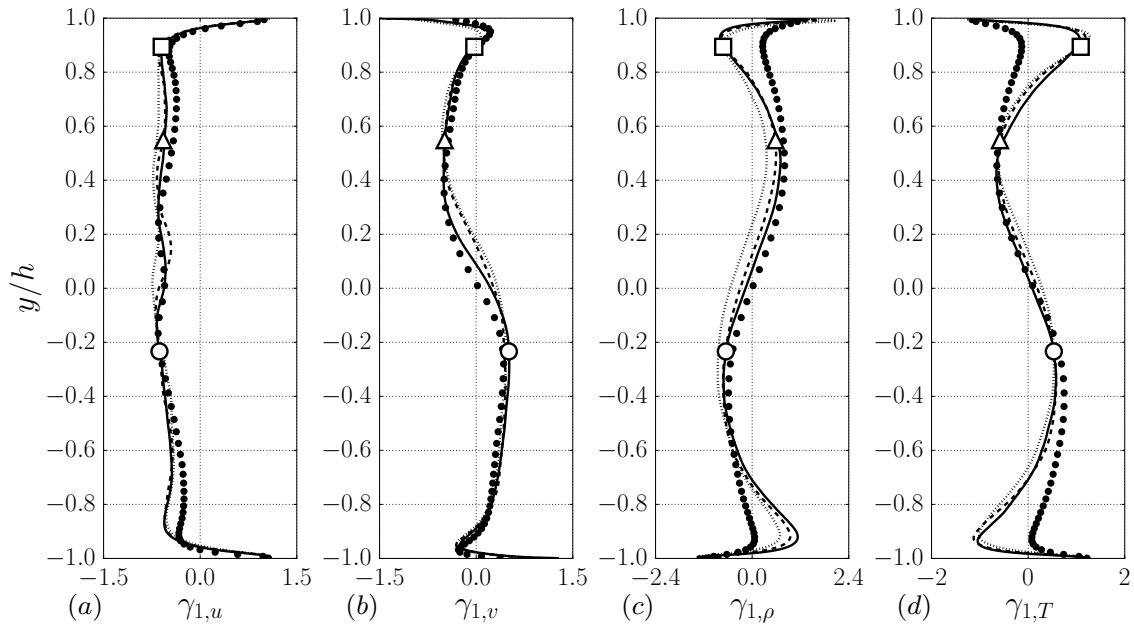

FIGURE 14. Skewness of the streamwise (*a*) and wall-normal (*b*) velocity component, density (*c*), and temperature (*d*) for $p_b = 1.1 p_{cr}$ and $\Delta T = 5$ K (——), 10 K (- - -), and 20 K ($\cdots$) and the ideal gas data (•). Average location of pseudotransition for $\Delta T=$ 5 K ($\circ$), 10 K ($\triangle$), and 20 K ($\square$).

where $\zeta = \frac{1}{2}\overline{\vec{\omega}'' \cdot \vec{\omega}''}$. While not as rigorous as the derivation based on Favre averaging outlined in Appendix C, the key advantage to (3.10) is that the right-hand side retains the same structure of (3.10), with the respective terms representing the same effects of the terms (*I*) to (*IV*). Also, for a large number of samples (i.e. grid points) in the direction of statistical homogeneity as it is the case for the finest grid resolution level adopted here (see table 2), one-time planar averages converge towards Reynolds averages. Mean profiles of the inviscid terms of (3.10) are shown in figure 13. As expected, the baroclinic fluctuating intensity increases with an increase of $\Delta T$. Simultaneously, the opposite trend is observed in the dilatational and stretching term (I) of the equation. The latter is the dominant term in the budgets.

Intuitively, one would expect an increase in vorticity, and concomitantly, of turbulence intensity in the presence of increasing density gradients, similar to the effect of increasing the density difference in a Rayleigh-Taylor instability. Yet the present results show a monotonic turbulence damping with increasing $\Delta T$, concentrated near the top wall of the channel (figure 8) on the pseudogaseous side, and is noted in all three velocity components (strongest in the wall-normal direction), and consistent with the vorticity statistics presented in this section.

## 4. High-Order Statistics, Probability Distribution Functions and Turbulent Spectra

The skewness of the fluctuating turbulent and thermodynamic quantities are presented in figure 14. The high-order moments of the velocity fluctuations are, for the most part, unaffected by real-fluid effects, although a more negative skewness of the streamwise velocity fluctuations is observed compared to ideal gas computations. The more noticeable difference is in the magnitude and sign of the skewness of density and temperature in the buffer layer regions; here the skewness of ideal gas density fluctuations approaches near zero values while for the real-fluid case it reaches an absolute maximum and with opposite sign with respect to the ideal gas case.

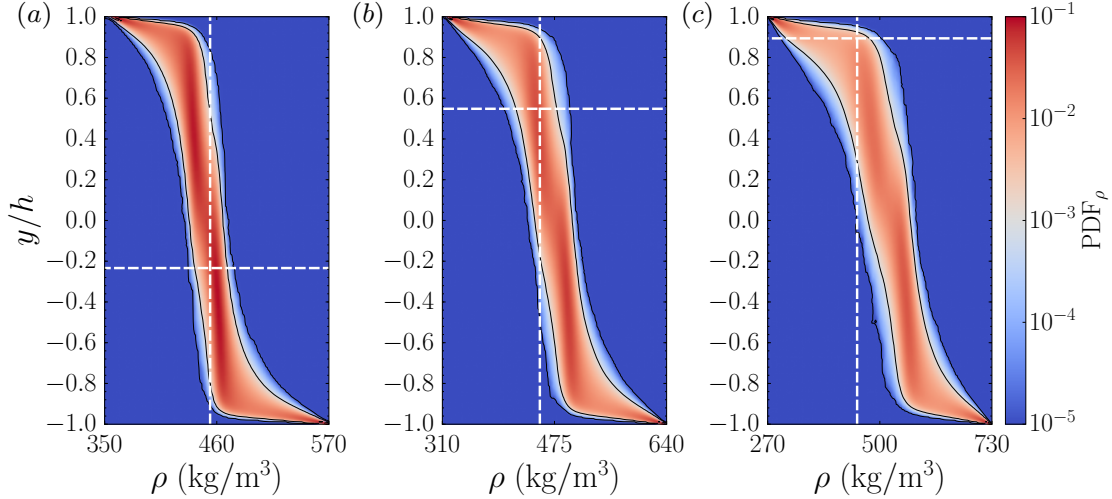

FIGURE 15. Contour of probability distribution function (PDF) of total density and the average location of pseudotransition, $y = y_{pb}$, pseudo-boiling density value $\rho = \rho_{pb}$ ($--$) for $p_b = 1.1 p_{cr}$ and $\Delta T = 5$ K ($a$), 10 K ($b$), and 20 K ($c$). The solid black line corresponds to the isocontour level $\text{PDF}_\rho = 10^{-3}$. Note that the plot extremes on the horizontal axis are increased for increasing $\Delta T$.

| $\Delta T$ (K) | $y = -0.97h$ | | | | $y = 0.97h$ | | | | $y = y_{pb}$ | | | |
|---|---|---|---|---|---|---|---|---|---|---|---|---|
| | $\rho'_{min}$ | $\rho'_{max}$ | $T'_{min}$ | $T'_{max}$ | $\rho'_{min}$ | $\rho'_{max}$ | $T'_{min}$ | $T'_{max}$ | $\rho'_{min}$ | $\rho'_{max}$ | $T'_{min}$ | $T'_{max}$ |
| | (kg/m$^3$) | | (K) | | (kg/m$^3$) | | (K) | | (kg/m$^3$) | | (K) | |
| 5  | $-49.58$  | $53.33$  | $-1.18$ | $1.13$ | $-37.76$ | $41.11$  | $-1.04$ | $1.24$ | $-23.36$  | $20.26$ | $-0.38$ | $0.44$ |
| 10 | $-67.58$  | $65.93$  | $-2.07$ | $1.56$ | $-59.18$ | $90.18$  | $-1.83$ | $2.51$ | $-42.69$  | $41.93$ | $-0.73$ | $0.83$ |
| 20 | $-101.55$ | $81.23$  | $-4.03$ | $3.34$ | $-91.68$ | $148.37$ | $-3.35$ | $5.39$ | $-124.41$ | $94.67$ | $-1.88$ | $4.06$ |

TABLE 5. Minimum and maximum values of fluctuating density and temperature at the approximate bottom-wall ($y \simeq -0.97h$) and top-wall ($y \simeq 0.97h$) rms peak locations and at the average location of pseudotransition, $y = y_{pb}$.

The positive peak in density skewness at the bottom wall is the result of intermittent events (discussed in more detail in section 5), which eject dense fluid from the pseudoliquid sublayer into the channel core kept in pseudoboiling conditions. Same considerations hold for the top wall, but in reverse, justifying the negative skewness peak of density observed there. No similar structure is observed in the skewness profiles of the ideal gas case. The skewness of temperature follows a specular pattern with respect to density, suggesting that fluctuations in pressure might not play a dominant role in the mass and momentum transport.

To gain more insight into the structure of thermodynamic fluctuations, probability density functions (PDF) of density and temperature have been extracted at all locations (figures 15–16). The PDFs widen as $\Delta T$ increases, as expected. Confirming previous observations, the largest variance is observed when pseudotransition takes place in the turbulent buffer layer, occurring at the top wall buffer-layer for $\Delta T = 20$ K. While the variance of the turbulent velocity fluctuations decreases with increasing $\Delta T$, the broader PDF of thermodynamic fluctuations is associated with the steepening of the corresponding gradients (figure 6).

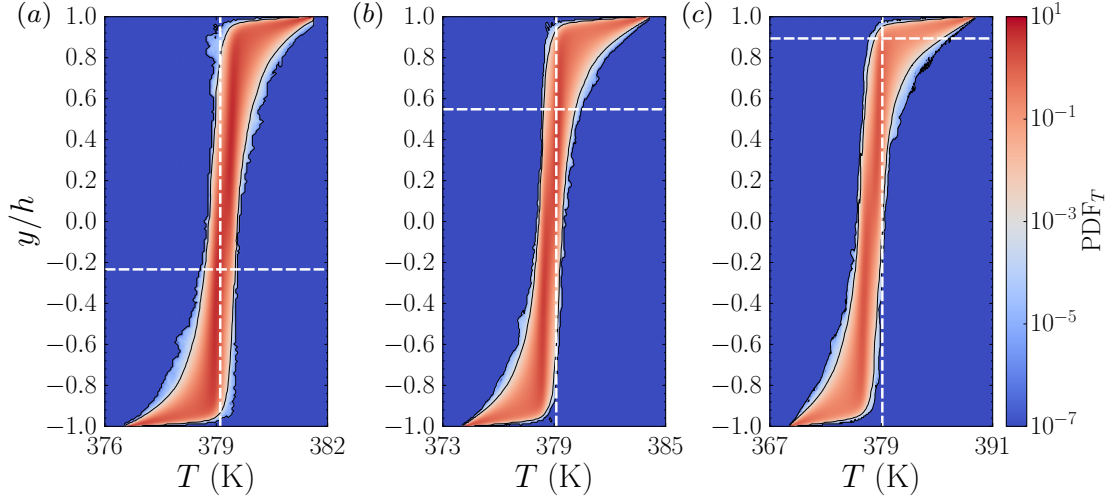

FIGURE 16. Contour of probability distribution function (PDF) of total temperature and the average location of pseudotransition, $y = y_{pb}$, pseudo-boiling density value $\rho = \rho_{pb}$ (- -) for $p_b = 1.1 p_{cr}$ and $\Delta T = 5$ K (*a*), 10 K (*b*), and 20 K (*c*). The solid black line corresponds to the isocontour level $\text{PDF}_\rho = 10^{-3}$. Note that the plot extremes on the horizontal axis are increased for increasing $\Delta T$.

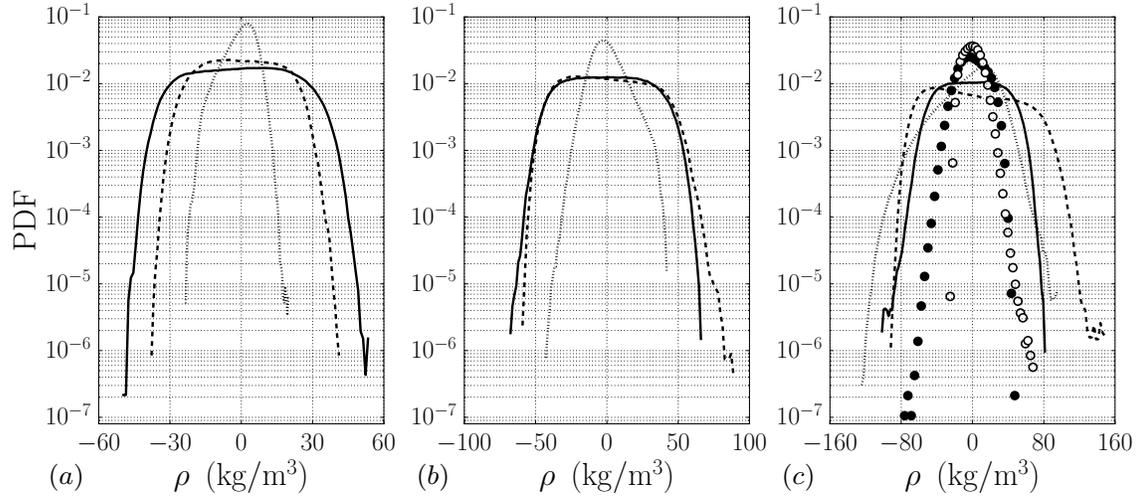

FIGURE 17. Probability distribution function (PDF) of fluctuating density at the bottom (——, ●) and top (- - -, ○) locations of peak $\rho_{rms}$ and at the average location of pseudotransition $y = y_{pb}$ (· · ·) for $p_b = 1.1 p_{cr}$ and $\Delta T = 5$ K (*a*), 10 K (*b*), and 20 K (*c*) with reference ideal gas data (circles). Note that the plot extremes on the horizontal axis are increased for increasing $\Delta T$.

The analysis in figures 17 and 18 focuses on three locations: the two buffer layers and the pseudophase transitioning location and includes a comparison with the ideal gas data. For both density and temperature, it is observed that the pseudo-phase transitioning region exhibits are much narrower distribution of the PDFs, whereas the buffer layers display a very pronounced kurtosis, with very rapid roll off at the tails of the distribution. Such PDF with very high kurtosis is not observed in the density PDF of the ideal gas case, while it is for the temperature PDF.

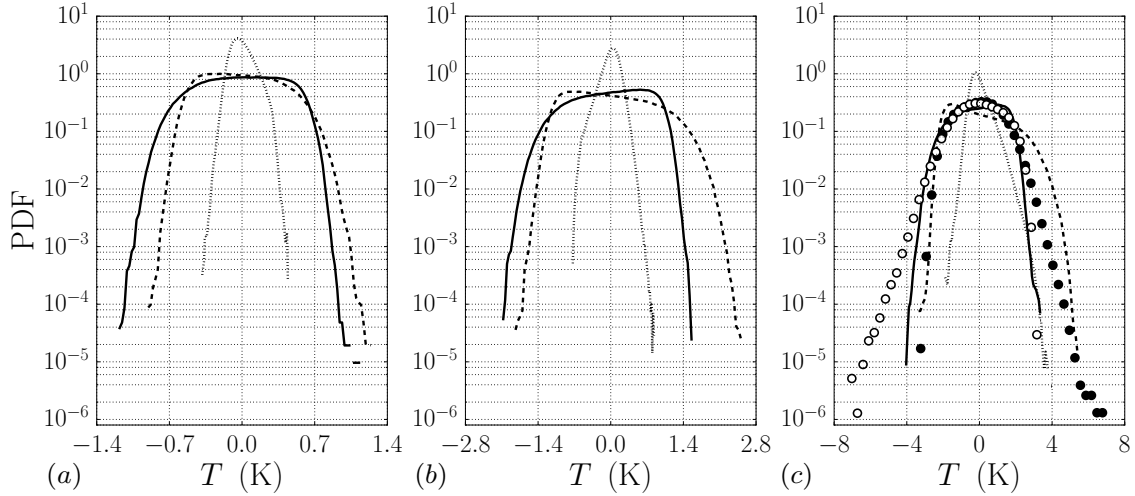

FIGURE 18. Probability distribution function (PDF) of fluctuating temperature at the bottom (———, ●) and top (- - -, ○) locations of peak $T_{rms}$ and at the average location of pseudotransition $y = y_{pb}$ (· · · ) for $p_b = 1.1 p_{cr}$ and $\Delta T = 5$ K (a), 10 K (b), and 20 K (c) with reference ideal gas data (circles). Note that the plot extremes on the horizontal axis are increased for increasing $\Delta T$.

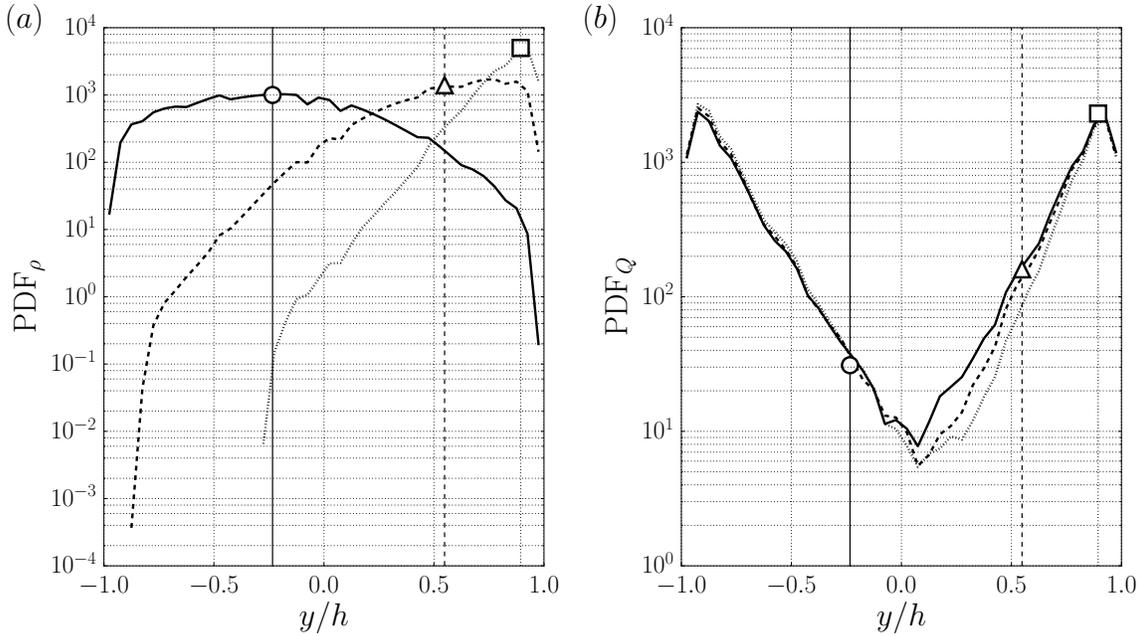

FIGURE 19. PDF of $y/h$ values conditional to $|\rho - \rho_{pb}| \leqslant 5.9$ kg/m$^3$ where $\rho_{pb} = 453.5$ kg/m$^3$ corresponding to $T_{pb} \pm 0.1$ K (a) and $Q = 2.49 \times 10^9$–$2.51 \times 10^9$ 1/s$^2$ (b) with average locations of the pseudotransition for $p_b = 1.1 p_{cr}$ and $\Delta T = 5$ K (———, ○), 10 K (- - -, △), and 20 K (· · · , □).

Figure 19a presents the PDF conditioned to a density range centered about its pseudotransitioning value. These results provide the probability of an instantaneous pseudophase change event at a given $y$ location, or, equivalently, the probability of the pseudointerface being instantaneously located at a given $y$ location. The location corresponding to the highest event count moves upwards in the channel as $\Delta T$ is increases and the distribution is narrowed; $y/h = -0.17$ for $\Delta T = 5$ K, 0.78 for $\Delta T = 10$ K, and 0.93 for $\Delta T = 20$ K. However, these values do not exactly match the average pseudotransition locations determined by the mean quantities (shown in figure 4 and indicated with symbols in figure 19), especially for $\Delta T = 10$ K. As a result, despite having a mean pseudotransition location at $y_{pb}/h = 0.55$ in the $\Delta T = 10$ K case, the greatest probability is much closer to the top wall, at about $y/h = 0.8 - 0.9$.

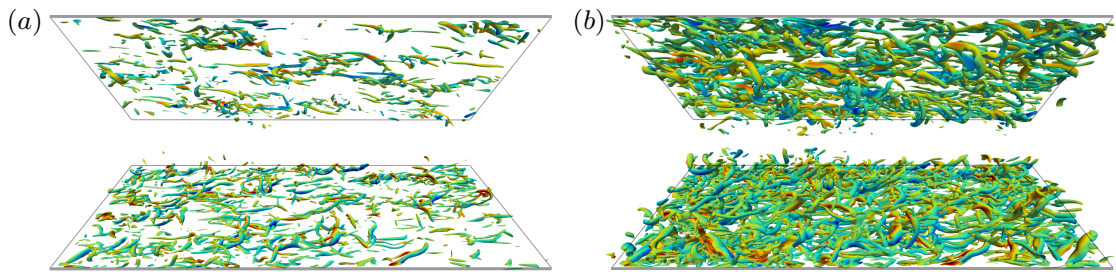

FIGURE 20. Isosurfaces of Q-criterion at $Q = 2.5 \times 10^9$ 1/s$^2$ (*a*) and $0.5 \times 10^9$ 1/s$^2$ (*b*) colored by the wall-normal velocity for $p_b = 1.1 p_{cr}$ and $\Delta T = 20$ K.

A coherent-structure-based probability distribution is also extracted (figure 19b). Conditioning the PDF on a selected Q-criteria value ($Q = 2.49 \times 10^9$–$2.51 \times 10^9$ 1/s$^2$, as used in figure 23), allows to identify the effects of increasing $\Delta T$ on the structural make-up of turbulence under transcritical conditions. In line with the observed turbulence damping in the vicinity of pseudotransition conditions, a reduction of the population density of turbulent structures in the top half of the channel is observed with increasing $\Delta T$. As quantitatively shown by the reduction in the number of observed events in the conditional statistics in (figure 19b). This effect is observed for several values of the Q-criterion (figure 20) and is consistent with the increasing asymmetry in the turbulent velocity profiles for increasing $\Delta T$ as shown in figure 8.

Figure 21 shows one-dimensional energy spectra of fluctuating density, wall-normal velocity, and temperature in the near-wall regions, which are heavily affected by the wall-generated turbulence. All the profiles roll off rapidly at high wavenumbers; providing further evidence of the adequacy of the resolution of both the hydrodynamic and thermodynamic quantities. The governing equations in conservative form, as expected, exhibit a build-up at high wavenumbers in the energy spectra of fluctuating pressure, which was mitigated by a high numerical resolution for the given Reynolds number (see Appendix B for grid convergence study).

The cospectrum of the wall-normal velocity and temperature fluctuations, $E_{vT}$, is also analyzed here to gain insight into the fundamental nature of their interaction. Its absolute value increases with $\Delta T$, as expected, given the increase of the wall-to-wall heat flux. Normalizing the cospectrum based on the single-variable spectra (figure 22) reveals an unexpected loss of transport efficiency, or coherence, at the pseudophase changing location for intermediate wave numbers as $\Delta T$ is increased; this is observed for both the streamwise and in the spanwise directions. Overall, the hydrodynamic and thermodynamic effects are highly correlated at or around the energy-containing turbulent length-scale.

## 5. Coherent Structures and Thermodynamics

Instantaneous isosurfaces of density and $Q$-criterion, as well as corresponding flooded contours of wall heat flux are shown in figure 23, for the bottom-wall only, to investigate the coupling between heat and mass transfer effects and the role coherent turbulent structures in the transport. The density isosurface at $\rho = 468$ kg/m$^3$ (value which corresponds to $y/h = -0.9$ in the mean density profile shown in figure 4) exhibits clear ejection events from the pseudoliquid region (near the cold, bottom wall) as the near-wall turbulence lifts-up the dense fluid into the lighter core of the channel. As the ejected fluid has more inertia than its lighter surrounding, and no gravitational effects are accounted for in these simulations, it reaches the core of the channel ($y/h = 0$, see red-colored surfaces) where the fluid undergoes a pseudophase change, effectively achieving mass transport. Naturally, the gravitational forces (in a stably stratified flow setup) would play a mitigating role in the observed mixing dynamics. This pseudophase change and the concomitant effects on the thermodynamics are a unique characteristic of transcritical flows and explain the high positive values of skewness of density (figure 14) in the bottom-half of the channel.

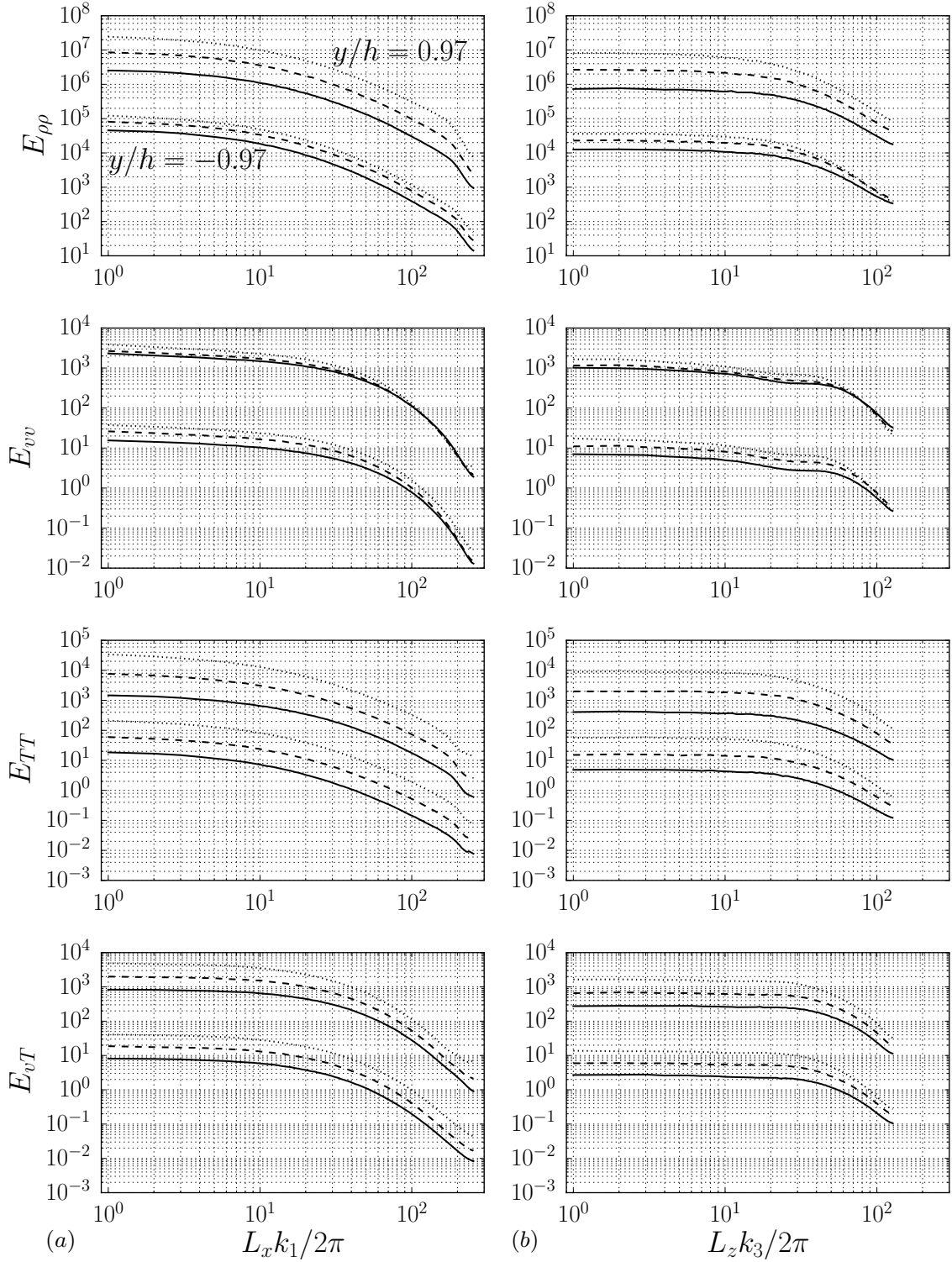

FIGURE 21. One-dimensional energy spectra of Reynolds averaged fluctuating density (first row), wall-normal velocity (second row), and temperature (third row) and one-dimensional cospectra between the Reynolds averaged fluctuating wall-normal velocity and temperature (fourth row) in the streamwise ($a$) and spanwise ($b$) direction extracted at the two near-wall peaks of density fluctuation intensity ($y/h = \pm 0.97$) for $p_b = 1.1 p_{cr}$ and $\Delta T = 5$ K (——), 10 K (- - -), and 20 K ($\cdots$). Spectra for the top wall data have been shifted vertically by 2 decades for clarity.

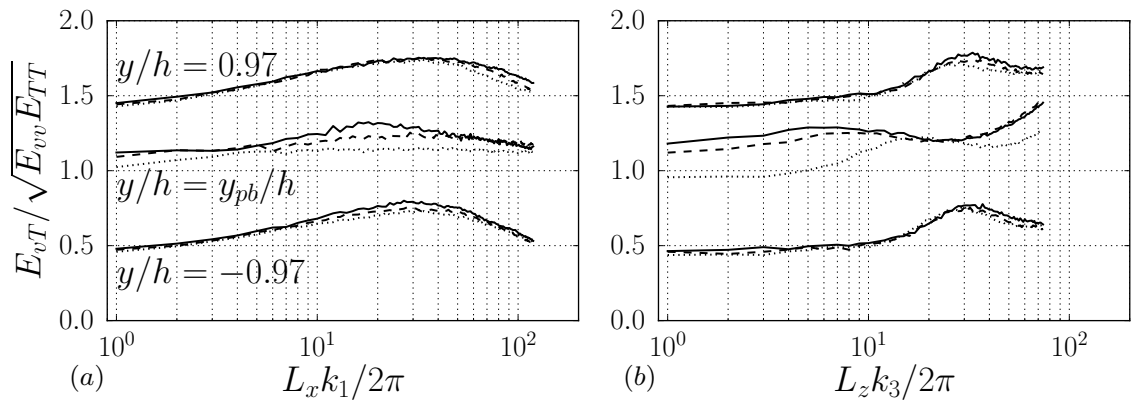

FIGURE 22. One-dimensional coherence between the Reynolds averaged fluctuating wall-normal velocity and temperature in the streamwise (*a*) and spanwise (*b*) direction extracted at the two near-wall peaks of density fluctuation intensity ($y/h = \pm 0.97$) and the average location of the pseudotransition based on the mean quantities for $p_b = 1.1 p_{cr}$ and $\Delta T = 5$ K (——), 10 K (- - -), and 20 K ($\cdots$). Coherence for the pseudotransition and the top near-wall peak data have been shifted vertically by 0.5 and 1.0 respectively for clarity.

The Q-criterion isosurface identifies the turbulent structures based on the velocity gradients alone. Interestingly, large-scale streamwise aligned structures are observed near the wall (see the circles), leading to the choice of a long computational domain length in the streamwise direction, 12 times the half-channel width, approximately twice the typical length required by the current friction Reynolds number (see table 2). Figure 23 (c) shows the corresponding elongated streaks in the wall-heat flux, spatially correlated with the ejection locations caused by the streamwise-elongated turbulent structures. Although, it is not too surprising that the dynamically active parts of the flow participate in the near wall heat transfer, the streamwise coherence.

Two-point velocity correlations in the streamwise and spanwise direction (figure 24) are extracted to confirm that, indeed, the computational box size has been adequately picked. A large streamwise and small spanwise coherence is observed near the top and bottom wall, confirming the visual observation of the narrow elongated streaks from figure 23. We note a much longer streamwise correlation length in the $u$ velocity (correlation reaches zero at about $0.15\ r_x/L_x$) than in $w$ (reaches zero at about $0.05\ r_x/L_x$). The lateral two-point correlation are consistent with the longitudinal ones and the three-dimensional visualizations. The signature of streamwise aligned streaks result in a short spanwise correlation length near the walls.

In the center of the channel, turbulence is nearly isotropic, a fact observed from turbulence statistics (figure 9) and from integral length scale analysis. The integral length scale (not shown) at the channel center is about 9% of the width. The integral length scales relative to the local Kolmogorov scale are presented in figure 25, once again, revealing the remarkably extended correlation length of the near wall structures.

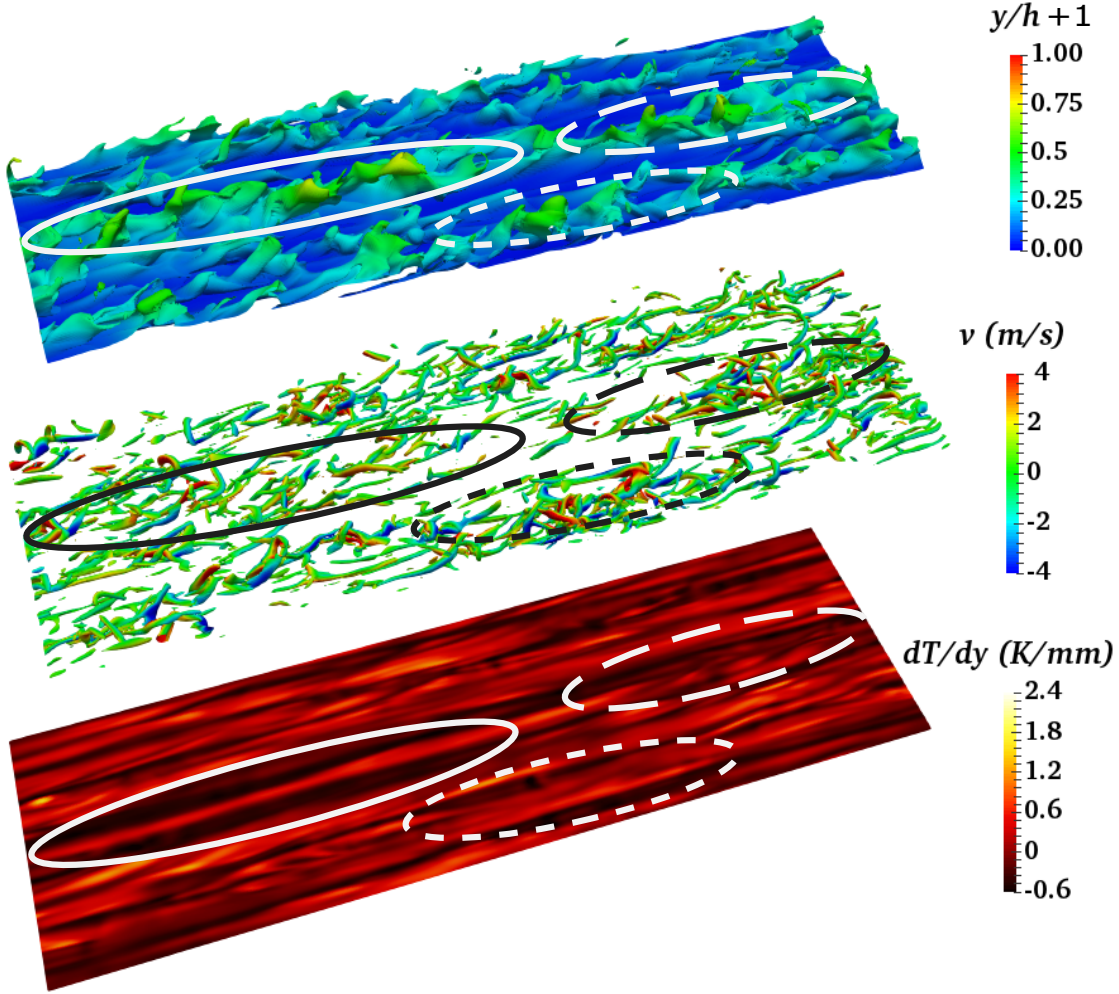

FIGURE 23. Isosurfaces of density ($\rho = 468$ kg/m$^3$) colored by the distance from the bottom wall (top), Q-criterion ($Q = 2.5 \times 10^9$ 1/s$^2$) colored by the wall-normal velocity (middle), and temperature gradient (bottom) for $p_b = 1.1 p_{cr}$ and $\Delta T = 5$ K *(enhanced online – see supplementary multimedia material)*.

In addition to the hydrodynamic correlations, thermodynamic two-point correlations are presented in figure 26. The two-point correlations for density and compressibility factor reflect the real-fluid characteristics discussed in figure 10. The correlations have the identical tendency showing the long streamwise and short spanwise correlation lengths near the walls and vice versa in the center region. The large streamwise coherence near the walls accords with the manifestation of pseudoliquid flow streaks observed in figure 23. The flow ejected from the walls in long streamwise streaks eventually take on a blob-like (shorter streamwise, longer spanwise structure) form as the ejected fluid reaches the channel centerplane. The strong similarity between all $\Delta T$ conditions is noted.

## 6. Conclusions

We have performed direct numerical simulations (DNS) of transcritical turbulent channel flow with differentially heated walls ($T_{top} - T_{bot} = \Delta T$) of R-134a (also called

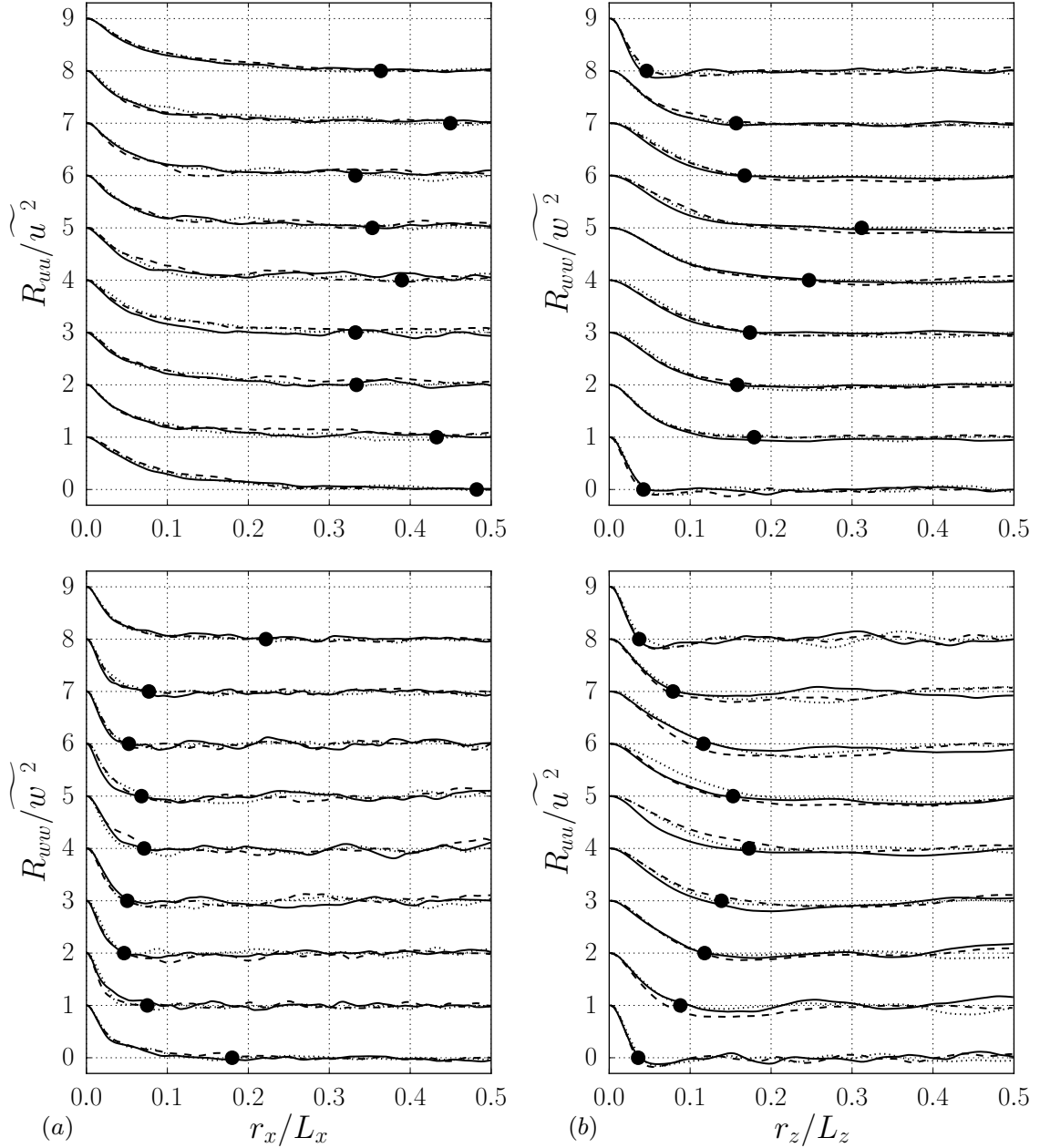

FIGURE 24. Normalized longitudinal (top) and lateral (bottom) two-point correlations of velocity in the streamwise (*a*) and spanwise (*b*) direction extracted at $y/h = -0.97, -0.75, -0.50, -0.25, 0.00, 0.25, 0.50, 0.75$, and $0.97$ for $p_b = 1.1 p_{cr}$ and $\Delta T = 5$ K (——), 10 K (- - -), and 20 K ($\cdots$). The lines have been shifted vertically corresponding to each $y/h$ from bottom to top. Average location of first zero-crossing points for $\Delta T = 5$ K, 10 K, and 20 K (●).

1,1,1,2-tetrafluoroethane, $CH_2FCF_3$) at a slightly supercritical pressure. By defining a statistically-steady turbulent channel flow at transcritical conditions, the turbulence and thermodynamic coupling could be studied. The simulations were conducted by solving the fully compressible Navier–Stokes equations using a conservative formulation. Special attention was paid to fully resolving all scales of the hydro- and thermodynamics of the setup to avoid non-physical oscillations which are characteristic of these flows. The Peng-Robinson (PR) equation of state was used with a consistent thermodynamic formulation to investigate the real-fluid effects. The simulations were run at a friction

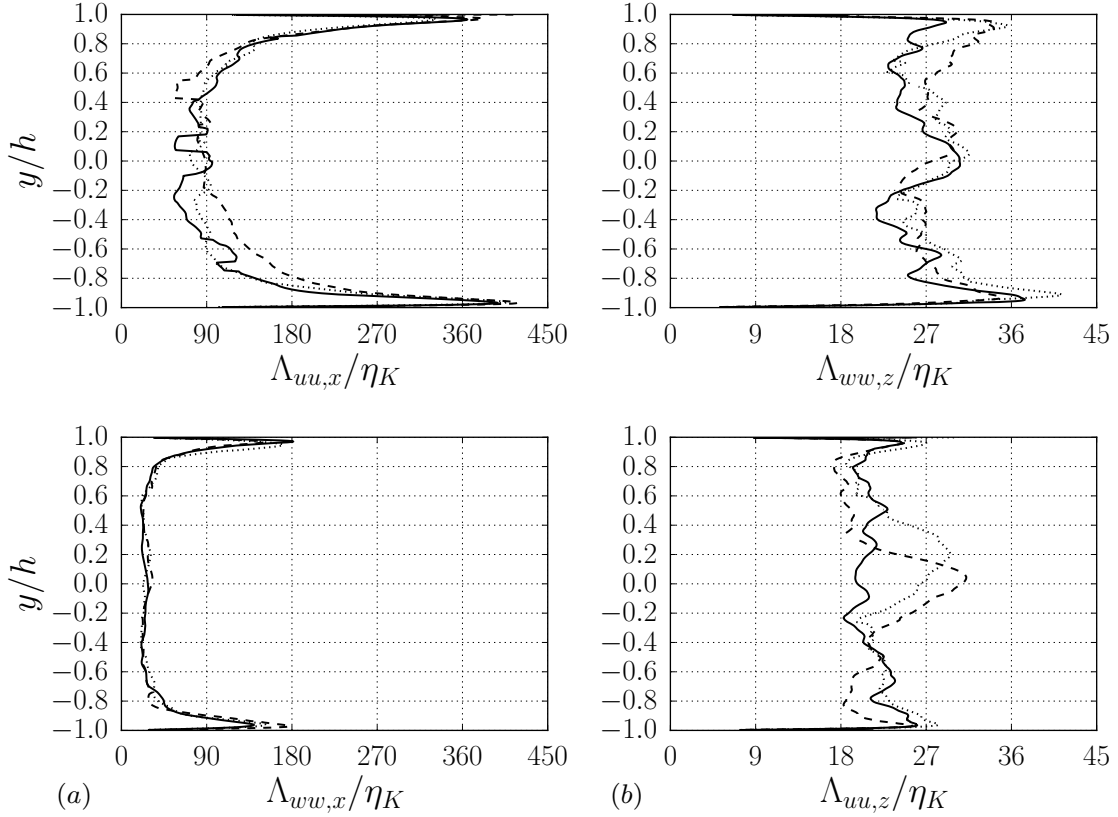

FIGURE 25. Ratio of the integral length scale (longitudinal (top) and lateral (bottom)) and the local Kolmogorov scale in the streamwise (a) and spanwise (b) direction. $\Delta T = 5$ K (——), 10 K (- - -), and 20 K ($\cdots$).

Reynolds numbers of about $Re_\tau = 373$. A realistic Prandtl number is used and computed from Chung's model to estimate the thermal conductivity. By varying the differential heating of the channel walls, the average location of the pseudophase change could be controlled, varying from $y/h = -0.23$, 0.55, and 0.89 for $\Delta T = 5$ K, 10 K, and 20 K, respectively. At the pseudophase change, the thermodynamic non-linearity are maximal, therefore the resulting effects of the thermodynamic non-linearities on turbulence could be investigated.

Conventional near-wall velocity scaling laws cannot capture the velocity distribution in transcritical flows due to the large density and thermophysical variations; even recent improvements to scaling laws for heated and cooled walls cannot accurately capture these effects. This leads us to conclude that additional wall modelling for transcritical flow is essential to correctly capture the near wall dynamics of transcritical flows. One justification for the near-wall modeling challenges stems from non-linear thermodynamic effects in the wall turbulence. The thermodynamics effects inhibit the turbulence through a decrease in the dilatational production term of the enstrophy equation. This term overpowers the increase of the baroclinic vorticity production. The profiles of the thermodynamic fluctuations show a higher intensity in the pseudogas (warm wall) compared to the pseudoliquid (cold wall) region; this occurs despite a reduction in the turbulence intensity near the top wall. The probability distribution function (PDF) of the thermodynamic fluctuations shows an broadening of the distribution with increasing differential heating. When the pseudophase change occurs near the wall ($\Delta T = 20$ K case), a highly skewed and very broad distribution is observed, which results from the nonlinear dynamics as the pseudophase change occurs near the viscous sub-layer. The structural signature of the turbulence in transcritical flows remains the most striking. Near the wall, the turbulence is aligned in long, yet meandering, streamwise coherent structures. The integral length of which are over 400 times the local Kolmogorov scale. The instantaneous visualizations and the two-point correlations have shown that strong ejections of heavy fluid into the channel core affect the structures and dynamics of turbulent channel flow, leave streaks in the temperature gradients at the wall.

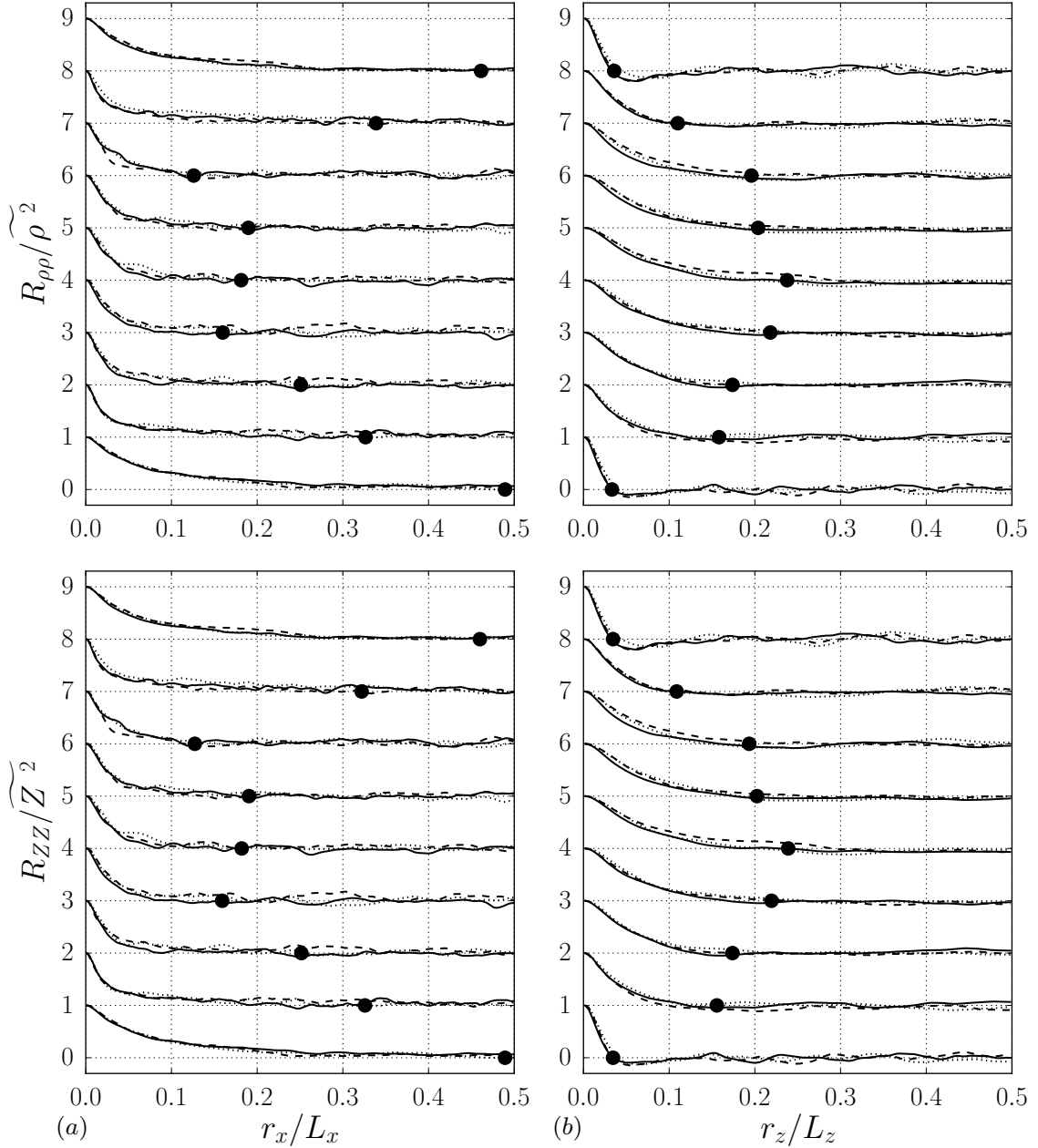

FIGURE 26. Normalized two-point correlations of density (top) and compressibility factor (bottom) in the streamwise (*a*) and spanwise (*b*) direction extracted at $y/h = -0.97, -0.75, -0.50, -0.25, 0.00, 0.25, 0.50, 0.75$, and $0.97$ for $p_b = 1.1 p_{cr}$ and $\Delta T = 5$ K (——), 10 K (- - -), and 20 K ($\cdots$). The lines have been shifted vertically corresponding to each $y/h$ from bottom to top. Average location of first zero-crossing points for $\Delta T$= 5 K, 10 K, and 20 K ($\bullet$).

## 7. Acknowledgements

This work has been performed under the support of the University Technology Center (UTC) in Advanced Thermal Management Systems at Purdue University lead by Mr. Pat Sweeney (Rolls-Royce, Indianapolis) and Prof. Stephen D. Heister (Purdue). The authors also thank Johan Larsson for providing us with the high-order structured code *Hybrid*. JPH acknowledges the support from the Natural Sciences and Engineering Research Council of Canada (NSERC) Discovery Grant program. The computing resources were provided by the Rosen Center for Advanced Computing (RCAC) at Purdue University and Information Technology at Purdue (ITaP).

## Appendix A. Modeling of Thermodynamic and Fluid Transport Properties

The PR equation of state reads

$$p = \frac{R_u T}{v_m - b} - \frac{a\alpha}{v_m^2 + 2bv_m - b^2} \tag{A 1}$$

$$a = \frac{0.45724 R_u^2 T_{cr}^2}{p_{cr}} \qquad b = \frac{0.07780 R_u T_{cr}}{p_{cr}}$$

$$\alpha = \left[1 + \left(0.37464 + 1.54226\omega - 0.26992\omega^2\right)\left(1 - T_r^{0.5}\right)\right]^2$$

where $R_u$ is the universal gas constant, $v_m$ the molar volume, $\omega$ the acentric factor, and $T_r = T/T_{cr}$ is the reduced temperature. The terms, $a$, $b$, and $\alpha$ account for intermolecular attractive and repulsive effects and the nonspherical shape of the molecules.

The thermodynamic relations based on the PR equation of state that incorporate departure functions are:

$$e(T, \rho) = e^0(T) + \frac{1}{\sqrt{8}bM_w}\left[T\left(\frac{\partial a\alpha}{\partial T}\right) - a\alpha\right]\ln\left(\frac{M_w + (1+\sqrt{2})b\rho}{M_w + (1-\sqrt{2})b\rho}\right) \tag{A 2}$$

$$h(T, \rho) = e(T, \rho) + \frac{p}{\rho} \tag{A 3}$$

$$c_v(T, \rho) = c_v^0(T) + \frac{T}{\sqrt{8}bM_w}\left(\frac{\partial^2 a\alpha}{\partial T^2}\right)\ln\left(\frac{M_w + (1+\sqrt{2})b\rho}{M_w + (1-\sqrt{2})b\rho}\right) \tag{A 4}$$

$$c_p(T, \rho) = c_v(T, \rho) + \frac{T}{\rho^2}\left(\frac{\partial p}{\partial T}\right)_\rho^2 \bigg/ \left(\frac{\partial p}{\partial \rho}\right)_T \tag{A 5}$$

$$\gamma(T, \rho) = \frac{c_p(T, \rho)}{c_v(T, \rho)} \tag{A 6}$$

$$c(T, \rho) = \sqrt{\gamma(T, \rho)\left(\frac{\partial p}{\partial \rho}\right)_T} \tag{A 7}$$

where $e$ is the internal energy, $h$ the enthalpy, $c_v$ the heat capacity at constant volume, $c_p$ the heat capacity at constant pressure, $\gamma$ the specific heat ratio, $c$ the speed of sound, and $M_w$ the molecular weight. The superscript, 0, denotes the thermodynamic property of the equivalent ideal gas state.

Departure functions derived from the selected equation of state assures full thermodynamic consistency (Sengers *et al.* 2000) of the simulations. As an example, here the partial derivatives in the relations for $c_p$ and $c$ are given by:

$$\left(\frac{\partial p}{\partial T}\right)_\rho = \frac{\rho R_u}{M_w - b\rho} - \left(\frac{\partial a\alpha}{\partial T}\right) \frac{\rho^2}{\left[M_w + \left(1+\sqrt{2}\right)b\rho\right]\left[M_w + \left(1-\sqrt{2}\right)b\rho\right]} \quad (A\,8)$$

$$\left(\frac{\partial p}{\partial \rho}\right)_T = \frac{M_w R_u T}{(M_w - b\rho)^2} - \frac{2a\alpha\rho M_w\left(M_w + b\rho\right)}{\left[M_w + \left(1+\sqrt{2}\right)b\rho\right]^2 \left[M_w + \left(1-\sqrt{2}\right)b\rho\right]^2} \quad (A\,9)$$

### A.1. *Thermophysical properties*

Chung's method (Chung *et al.* 1988) is used to obtain transport properties such as viscosity and thermal conductivity. The viscosity is given by

$$\mu = \mu^* \frac{36.344\left(M_w T_{cr}\right)^{1/2}}{v_{m,c}^{2/3}} \quad (A\,10)$$

where $v_{m,c}$ is the critical molar volume and $\mu^*$ is

$$\mu^* = \frac{(T^*)^{1/2}}{\Omega_v} F_c \left[(G_2)^{-1} + A_6 y\right] + \mu^{**} \quad (A\,11)$$

$T^*$, $\Omega_v$, and $F_c$ are given as

$$T^* = 1.2593 T_r \quad (A\,12)$$

$$\Omega_v = \left[A\left(T^*\right)^{-B}\right] + C\left[\exp\left(-DT^*\right)\right] + E\left[\exp\left(-FT^*\right)\right] + GT^{*B}\sin\left(ST^{*W} - H\right) \quad (A\,13)$$

$$F_c = 1 - 0.2756\omega + 0.059035\mu_r^4 + \kappa_a \quad (A\,14)$$

where $\kappa_a$ is the association factor for hydrogen bonding effect of highly polar substances such as alcohols and acids, $\Omega_v$ and $F_c$ mean the viscosity collision integral and consideration for the shape and polarity of molecules for dilute gases, respectively. The dimensionless dipole moment, $\mu_r$, is given by

$$\mu_r = 131.3 \frac{\chi}{\left(v_{m,c} T_{cr}\right)^{1/2}} \quad (A\,15)$$

where $\chi$ is the dipole moment of molecule.

The other terms appearing in the relationships above are as follows:

$$y = \frac{\rho v_{m,c}}{6} \quad (A\,16)$$

$$G_1 = \frac{1 - 0.5y}{(1-y)^3} \quad (A\,17)$$

$$G_2 = \frac{A_1\left[\left[1 - exp\left(-A_4 y\right)\right]/y\right] + A_2 G_1 exp\left(A_5 y\right) + A_3 G_1}{A_1 A_4 + A_2 + A_3} \quad (A\,18)$$

$$\mu^{**} = A_7 y^2 G_2 exp\left[A_8 + A_9\left(T^*\right)^{-1} + A_{10}\left(T^*\right)^{-2}\right] \quad (A\,19)$$

$$A_i = a_0\left(i\right) + a_1\left(i\right)\omega + a_2\left(i\right)\mu_r^4 + a_3\left(i\right)\kappa_a \quad (A\,20)$$

The thermal conductivity was developed by following a similar approach to viscosity.

$$\lambda = \frac{31.2\mu^0 \Psi}{M'_w}\left(G_2^{-1} + B_6 y\right) + qB_7 y^2 T_r^{1/2} G_2 \tag{A 21}$$

where

$$\mu^0 = 40.785\frac{F_c (M_w T)^{1/2}}{v_{m,c}^{2/3} \Omega_v} \tag{A 22}$$

$$\Psi = 1 + \alpha\left(\frac{0.215 + 0.28288\alpha - 1.061\beta + 0.26665Z}{0.6366 + \beta Z + 1.061\alpha\beta}\right) \tag{A 23}$$

$$\alpha = \frac{c_v}{R_u} - 1.5 \tag{A 24}$$

$$\beta = 0.7862 - 0.7109\omega + 1.3168\omega^2 \tag{A 25}$$

$$Z = 2.0 + 10.5 T_r^2 \tag{A 26}$$

$$M'_w = M_w/10^3 \tag{A 27}$$

$$q = 3.586 \times 10^{-3} \frac{(T_{cr}/M'_w)^{1/2}}{v_{m,c}^{2/3}} \tag{A 28}$$

For the term, $G_2$, the form is identical to the one of viscosity, but $A_i$ is replaced with $B_i$ which has the different values. All the other terms that are not defined and the empirical coefficients are found in Poling *et al.* (2001).

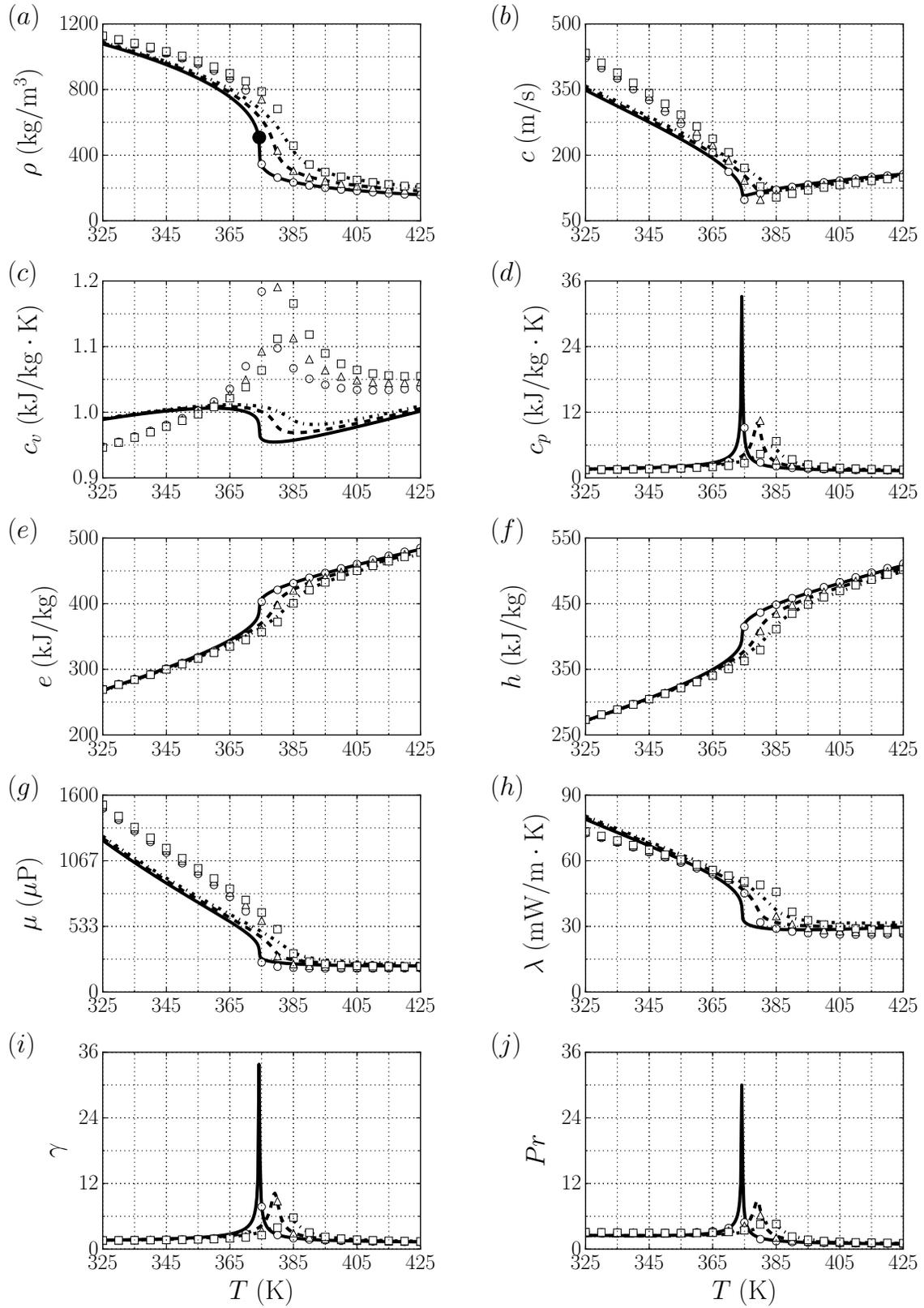

FIGURE 27. Thermodynamic properties predicted by the PR equation of state and the Chung's model (lines) and the NIST data (symbols) at various pressure conditions (——— ○, 40.590 bar; - - - △, 44.649 bar; -·· □, 48.708 bar). (*a*) Density (● critical point). (*b*) Speed of sound. (*c*) Heat capacity at constant volume. (*d*) Heat capacity at constant pressure. (*e*) Internal energy. (*f*) Enthalpy. (*g*) Dynamic viscosity. (*h*) Thermal conductivity. (*i*) Specific heat ratio. (*j*) Prandtl number.

| $N_x \times N_y \times N_z$ | | $64\times96\times64$ | $128\times128\times96$ | $192\times128\times128$ | $384\times256\times256$ | $512\times256\times256$ |
|---|---|---|---|---|---|---|
| Index | $i, j,$ or $k-1$ | 0.00005 | 0.00015 | 0.00025 | 0.00045 | 0.00055 |
| | $i, j,$ or $k$ | 0.99990 | 0.99970 | 0.99950 | 0.99910 | 0.99890 |
| | $i, j,$ or $k+1$ | 0.00005 | 0.00015 | 0.00025 | 0.00045 | 0.00055 |

TABLE 6. Filtering factors used in the top-hat filter

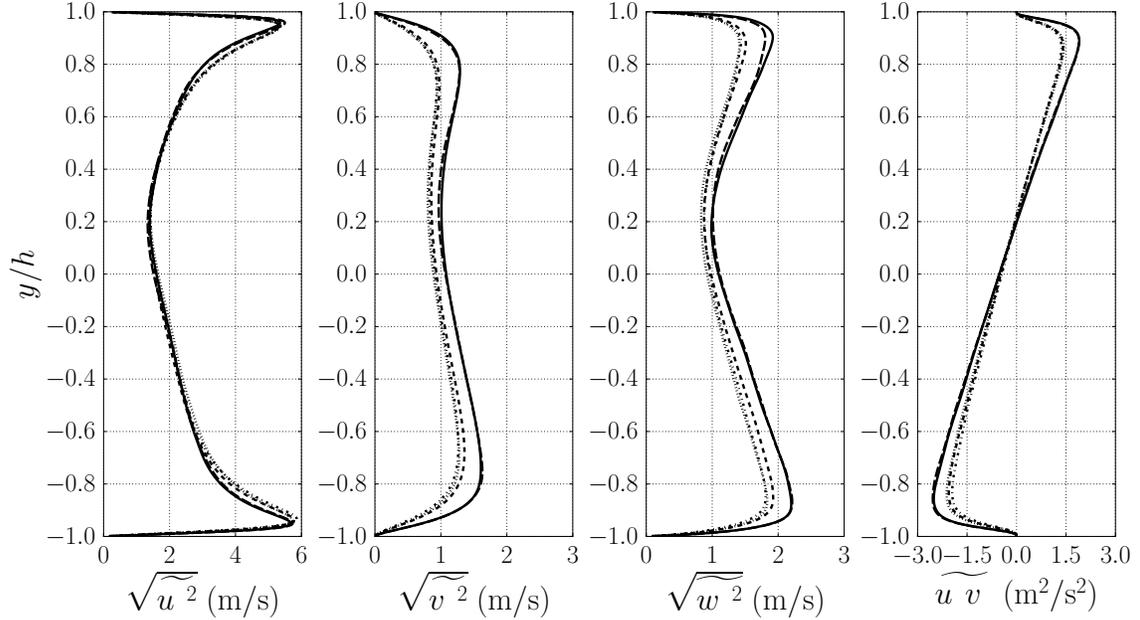

FIGURE 28. RMS of the Favre averaged fluctuating velocities, $u''$, $v''$, and $w''$, and the Reynolds shear stress by $u''$ and $v''$ for $p_b = 1.1 p_{cr}$ and $\Delta T = 20$ K at grid resolution of $64\times96\times64$ ($\cdots$), $128\times128\times96$ (-$\cdot\cdot$-), $192\times128\times128$ (- - -), $384\times256\times256$ (– –), and $512\times256\times256$ (——).

## Appendix B. Grid Convergence Study

Grid convergence of transcritical flows is essential to determine the adequacy of a direct numerical simulation as we recall that the minimal thermodynamic length scale to be resolved in transcritical flows is typically smaller than the Kolmogorov length scale. Insufficient spatial resolution is typically evidenced by a large spectral pile-up in the thermodynamic quantities; in which case, the obtained results should be considered erroneous. Here, the grid sensitivity is investigated for the most critical case of $\Delta T$ = 20 K. Figure 28 shows the grid sensitivity of the velocity RMS. We highlight the insensitivity of the streamwise fluctuations to the grid resolution, whereas an unresolved simulation underestimates the peak fluctuations in the spanwise and wall normal velocity components. The overall trends of the RMS profiles (asymmetry, relative peak height etc.) are independent of the grid resolution.

The grid sensitivity of thermodynamic fluctuations is shown in figures 29 and 30. We note a slow convergence of the thermodynamic quantities, particularly for the pressure. Figure 30 shows that the fluctuating enthalpy RMS is well captured on a coarse mesh. But the turbulent enthalpy flux, an important quantity for the characterization of the convective heat transfer, requires a large grid count for a correct estimation. An insufficient grid resolution will underestimate the magnitude of the turbulence effect on the heat transfer in this transcritical system.

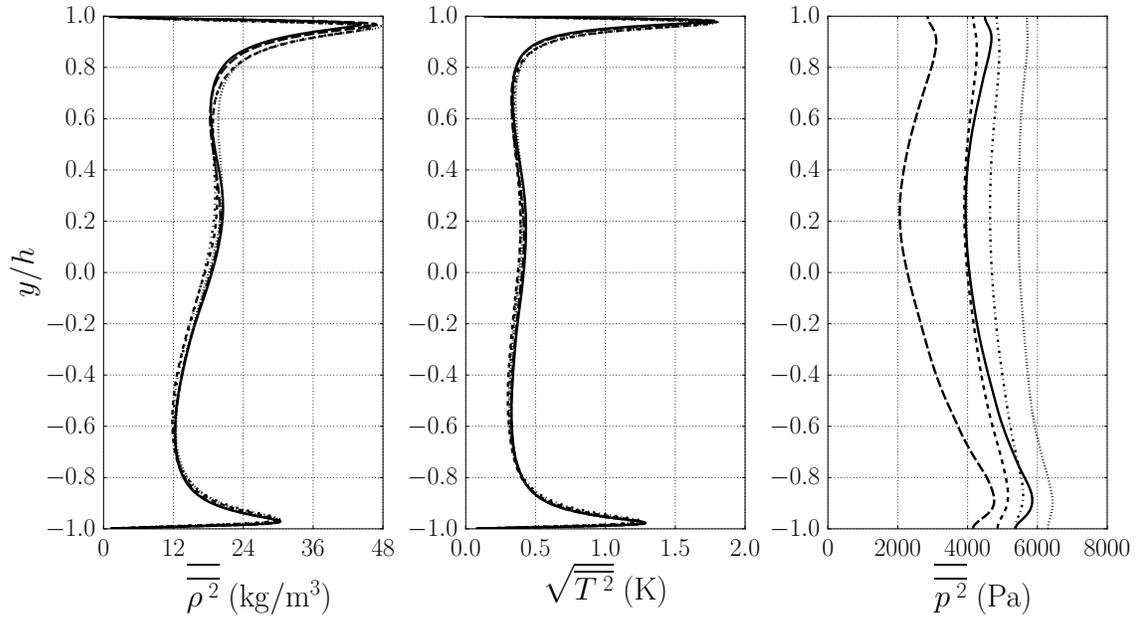

FIGURE 29. Root-mean-square of Reynolds density, temperature, and pressure fluctuations for $p_b$ = $1.1p_{cr}$ and $\Delta T$ = 20 K at grid resolution of 64×96×64 (· · · ), 128×128×96 (-·-), 192×128×128 (- - -), 384×256×256 (– –), and 512×256×256 (——).

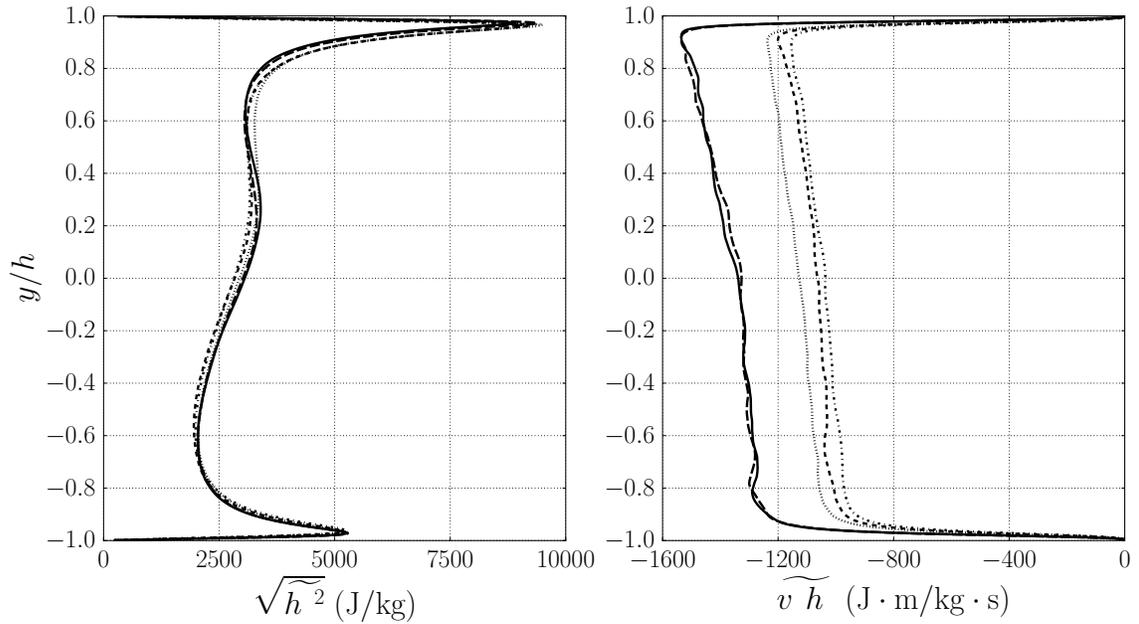

FIGURE 30. RMS of the Favre averaged fluctuating enthalpy and wall-normal turbulent enthalpy flux for $p_b$ = $1.1p_{cr}$ and $\Delta T$ = 20 K at grid resolution of 64×96×64 (· · · ), 128×128×96 (-·-), 192×128×128 (- - -), 384×256×256 (– –), and 512×256×256 (——).

The one-dimensional energy spectra of fluctuating density, wall-normal velocity, and pressure in the streamwise and spanwise directions are presented in figure 31. The profiles are extracted at $y/h$ = $-0.97$, 0, and 0.97 which correspond to the location of the

34                        K. Kim, J.-P. Hickey and C. Scalo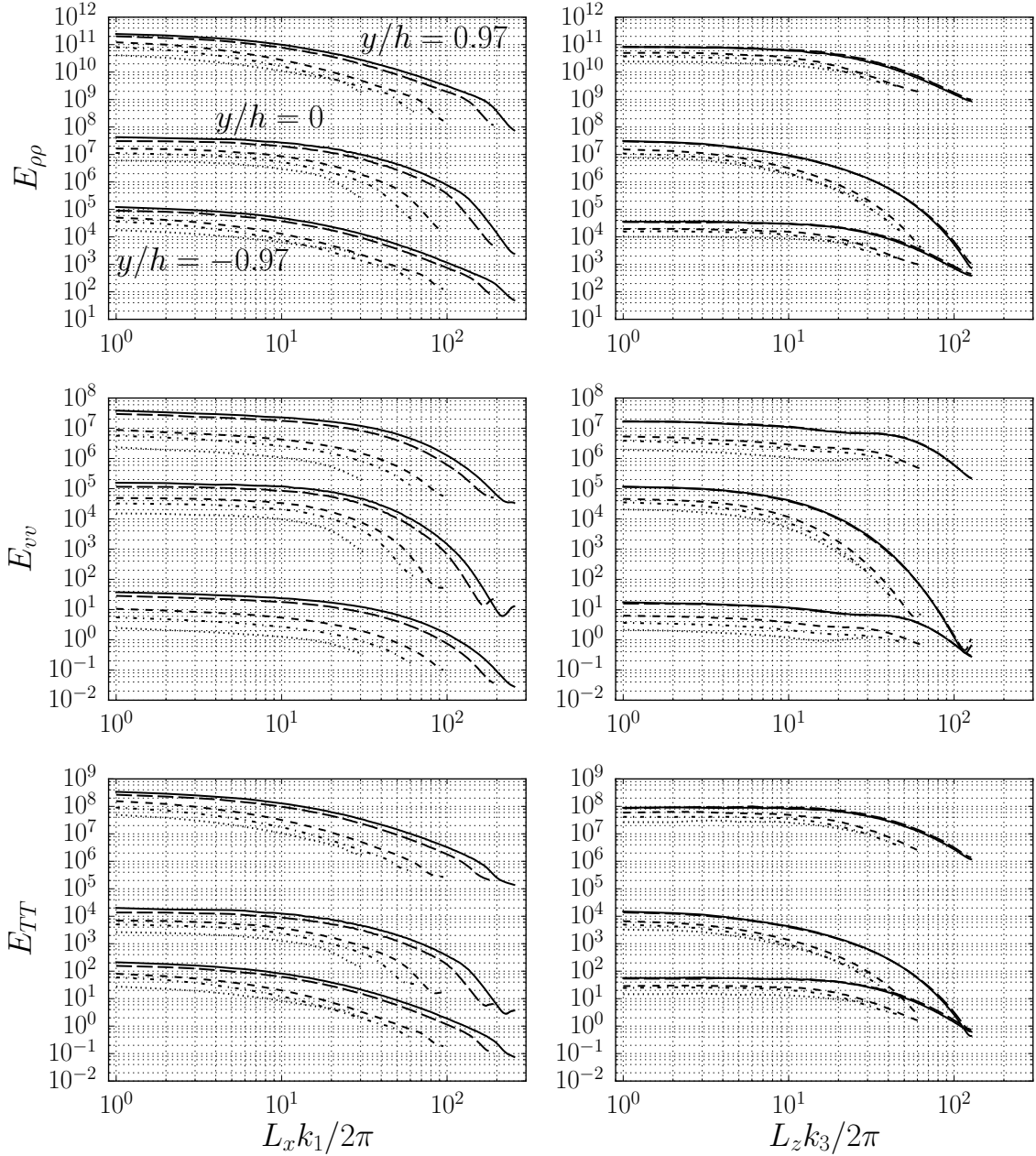

FIGURE 31. One-dimensional energy spectra of Reynolds averaged fluctuating density (top), wall-normal velocity (middle), and temperature (bottom) in the streamwise (*a*) and spanwise (*b*) directions extracted at the two near-wall peaks of density fluctuation intensity ($y/h = \pm 0.97$) and the centerline ($y/h = 0$) for $p_b = 1.1 p_{cr}$ and $\Delta T = 20$ K at grid resolution of $64 \times 96 \times 64$ (···), $128 \times 128 \times 96$ (-··-), $192 \times 128 \times 128$ (- - -), $384 \times 256 \times 256$ (– –), and $512 \times 256 \times 256$ (——). Spectra for the centerline and the top wall data have been shifted vertically by 3 decades and 6 decades respectively for clarity.

thermodynamic RMS peaks (top and bottom wall) and the centerplane. As the grid resolution increases, the spectral broadening is observed with a slight increase at the high wavenumbers.

Figure 32 presents the average profiles of the normalized Kolmogorov length scale in the streamwise, wall-normal, and spanwise directions. The Kolmogorov scale, $\eta_K$, which

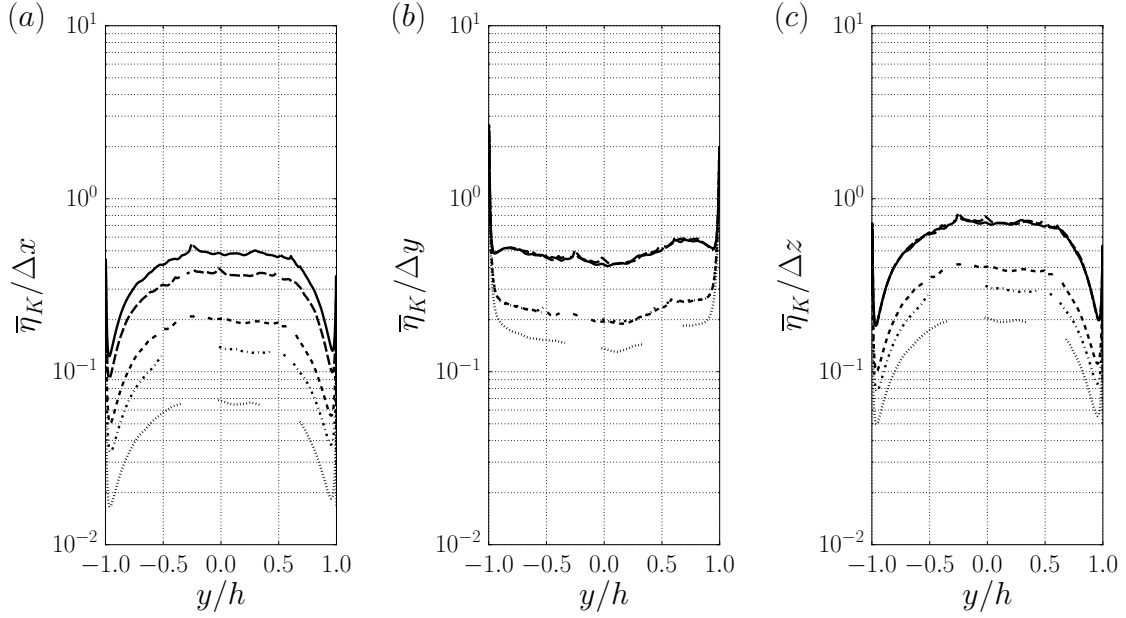

FIGURE 32. Normalized average Kolmogorov length scales, $\overline{\eta}_K/\Delta x$ (a), $\overline{\eta}_K/\Delta y$ (b), and $\overline{\eta}_K/\Delta z$ (c), at $p_b = 1.1 p_{cr}$ and $\Delta T = 20$ K at grid resolution of 64×96×64 (· · ·), 128×128×96 (- · -), 192×128×128 (- - -), 384×256×256 (– –), and 512×256×256 (——).

quantifies the smallest turbulence length scale, is defined as :

$$\eta_K \equiv \left(\frac{\nu^3}{\epsilon}\right)^{1/4} \tag{B 1}$$

where $\nu$ and $\epsilon$ represent kinematic viscosity and dissipation rate of TKE per unit mass. For compressible flows $\epsilon$ is defined as:

$$\epsilon \equiv \frac{1}{\overline{\rho}}\overline{\tau_{ij}\frac{\partial u_i''}{\partial x_j}} \tag{B 2}$$

The profiles of the normalized Kolmogorov length scale approach unity as the grid resolution increases. It is observed that the flow in liquid-like phase needs a finer grid than that in the gas-like phase to resolve the turbulence length scale, which is a result of the larger density at the cooled wall. These figures show adequate grid resolution in the wall-normal direction, especially near the walls. This study also highlighted the importance of a sufficient resolution in the streamwise direction as well. This imposed severe constraints on the resolution.

## Appendix C. Enstrophy Budget Transport Equation

### C.1. *Total Vorticity Budgets*

The first step is to break down the various contributions in the total vorticity transport equation to derive the enstrophy budget transport equation. The non-averaged total vorticity budget reads,

$$\frac{\partial \omega_i}{\partial t} + u_j \frac{\partial \omega_i}{\partial x_j} = \underbrace{\omega_j \frac{\partial u_i}{\partial x_j}}_{I} - \underbrace{\omega_i \frac{\partial u_j}{\partial x_j}}_{II} + \underbrace{\epsilon_{ijk} \frac{1}{\rho^2} \frac{\partial \rho}{\partial x_j} \frac{\partial p}{\partial x_k}}_{III} + \underbrace{\epsilon_{ijk} \frac{\partial}{\partial x_j} \left( \frac{1}{\rho} \frac{\partial \tau_{km}}{\partial x_m} \right)}_{IV} \quad (C\,1)$$

where each term describes vorticity stretching and tilting by the flow velocity gradients (I), vorticity stretching by the dilatational field (II), baroclinic effects (III), vorticity diffusion by the viscous effect (IV).

### C.2. *Favre-Averaged Total Vorticity*

First, we derive the $\rho$-scaled total vorticity transport equation:

$$\rho \frac{\partial \omega_i}{\partial t} + \rho u_j \frac{\partial \omega_i}{\partial x_j} = \rho \omega_j \frac{\partial u_i}{\partial x_j} - \rho \omega_i \frac{\partial u_j}{\partial x_j} + \rho \epsilon_{ijk} \frac{1}{\rho^2} \frac{\partial \rho}{\partial x_j} \frac{\partial p}{\partial x_k} + \rho \epsilon_{ijk} \frac{\partial}{\partial x_j} \left( \frac{1}{\rho} \frac{\partial \tau_{km}}{\partial x_m} \right) \quad (C\,2)$$

and $\omega_i$-scaled continuity equation:

$$\omega_i \frac{\partial \rho}{\partial t} + \omega_i \frac{\partial \rho u_j}{\partial x_j} = \omega_i \frac{\partial \rho}{\partial t} + \frac{\partial \rho \omega_i u_j}{\partial x_j} - \rho u_j \frac{\partial \omega_i}{\partial x_j} = 0 \quad (C\,3)$$

This produces a seemingly conservative version of the vorticity equation. Recombining the two equations yields

$$\begin{aligned}
\frac{\partial \rho \omega_i}{\partial t} + \frac{\partial \rho \omega_i u_j}{\partial x_j} - \rho u_j \frac{\partial \omega_i}{\partial x_j} + \rho u_j \frac{\partial \omega_i}{\partial x_j} = \\
\rho \omega_j \frac{\partial u_i}{\partial x_j} - \rho \omega_i \frac{\partial u_j}{\partial x_j} + \rho \epsilon_{ijk} \frac{1}{\rho^2} \frac{\partial \rho}{\partial x_j} \frac{\partial p}{\partial x_k} + \rho \epsilon_{ijk} \frac{\partial}{\partial x_j} \left( \frac{1}{\rho} \frac{\partial \tau_{km}}{\partial x_m} \right)
\end{aligned} \quad (C\,4)$$

Introducing the Reynolds-averaging yields

$$\begin{aligned}
\overline{\frac{\partial \rho \omega_i}{\partial t}} + \overline{\frac{\partial \rho \omega_i u_j}{\partial x_j}} - \overline{\rho u_j \frac{\partial \omega_i}{\partial x_j}} + \overline{\rho u_j \frac{\partial \omega_i}{\partial x_j}} = \\
\overline{\rho \omega_j \frac{\partial u_i}{\partial x_j}} - \overline{\rho \omega_i \frac{\partial u_j}{\partial x_j}} + \overline{\rho \epsilon_{ijk} \frac{1}{\rho^2} \frac{\partial \rho}{\partial x_j} \frac{\partial p}{\partial x_k}} + \overline{\rho \epsilon_{ijk} \frac{\partial}{\partial x_j} \left( \frac{1}{\rho} \frac{\partial \tau_{km}}{\partial x_m} \right)}
\end{aligned} \quad (C\,5)$$

and by definition of Favre-averaging $\widetilde{(\cdot)}$,

$$\overline{\rho} \widetilde{u} = \overline{\rho u} \quad (C\,6)$$

and assuming commutativity of the averaging operation with the derivative yields

$$\begin{aligned}
\frac{\partial \overline{\rho} \widetilde{\omega}_i}{\partial t} + \frac{\partial \overline{\rho} \widetilde{\omega_i u_j}}{\partial x_j} - \overline{\rho} \widetilde{u_j \frac{\partial \omega_i}{\partial x_j}} + \overline{\rho} \widetilde{u_j \frac{\partial \omega_i}{\partial x_j}} = \\
\overline{\rho} \widetilde{\omega_j \frac{\partial u_i}{\partial x_j}} - \overline{\rho} \widetilde{\omega_i \frac{\partial u_j}{\partial x_j}} + \overline{\rho} \epsilon_{ijk} \frac{1}{\rho^2} \widetilde{\frac{\partial \rho}{\partial x_j} \frac{\partial p}{\partial x_k}} + \overline{\rho} \epsilon_{ijk} \frac{\partial}{\partial x_j} \widetilde{\left( \frac{1}{\rho} \frac{\partial \tau_{km}}{\partial x_m} \right)}
\end{aligned} \quad (C\,7)$$

Taking the Reynolds-averaged continuity equation scaled by $\widetilde{\omega}_i$:

$$\widetilde{\omega}_i \frac{\partial \overline{\rho}}{\partial t} + \widetilde{\omega}_i \frac{\partial \overline{\rho u_j}}{\partial x_j} = \widetilde{\omega}_i \frac{\partial \overline{\rho}}{\partial t} + \frac{\partial \overline{\rho} \widetilde{\omega}_i \widetilde{u}_j}{\partial x_j} - \overline{\rho} \widetilde{u}_j \frac{\partial \widetilde{\omega}_i}{\partial x_j} = 0 \quad (C\,8)$$

and subtracting (C 8) from (C 7) and dividing by $\bar{\rho}$ gives

$$\frac{\partial \widetilde{\omega_i}}{\partial t} + \frac{1}{\bar{\rho}}\frac{\partial \overline{\rho\widetilde{\omega_i u_j}}}{\partial x_j} - \widetilde{u_j \frac{\partial \omega_i}{\partial x_j}} - \frac{1}{\bar{\rho}}\frac{\partial \overline{\rho\widetilde{\omega_i}\widetilde{u_j}}}{\partial x_j} + \widetilde{u_j}\frac{\partial \widetilde{\omega_i}}{\partial x_j} + \widetilde{u_j\frac{\partial \omega_i}{\partial x_j}} = \\ \widetilde{\omega_j \frac{\partial u_i}{\partial x_j}} - \widetilde{\omega_i \frac{\partial u_j}{\partial x_j}} + \epsilon_{ijk}\widetilde{\frac{1}{\rho^2}\frac{\partial \rho}{\partial x_j}\frac{\partial p}{\partial x_k}} + \epsilon_{ijk}\widetilde{\frac{\partial}{\partial x_j}\left(\frac{1}{\rho}\frac{\partial \tau_{km}}{\partial x_m}\right)} \quad (C\,9)$$

### C.3. *Favre Fluctuations of Vorticity*

We define enstrophy of the fluctuating field as

$$\xi'' = \omega_l'' \omega_l'' \quad (C\,10)$$

It is straightforward enough to derive a transport of the fluctuations of vorticity $\omega_i'' = \omega_i - \widetilde{\omega_i}$. Subtracting the previous equation (C 9) from the original nonconservative total vorticity equation (C 1) and then multiplying $\omega_i''$ yields

$$\frac{1}{2}\frac{\partial \omega_i'' \omega_i''}{\partial t} - \omega_i''\frac{1}{\bar{\rho}}\frac{\partial \overline{\rho\widetilde{\omega_i u_j}}}{\partial x_j} + \omega_i''\frac{1}{\bar{\rho}}\frac{\partial \overline{\rho\widetilde{\omega_i}\widetilde{u_j}}}{\partial x_j} - \omega_i''\widetilde{u_j}\frac{\partial \widetilde{\omega_i}}{\partial x_j} + \omega_i'' u_j\frac{\partial \omega_i}{\partial x_j} = \\ \omega_i''\left(\omega_j\frac{\partial u_i}{\partial x_j}\right)'' - \omega_i''\left(\omega_i\frac{\partial u_j}{\partial x_j}\right)'' + \omega_i''\epsilon_{ijk}\left(\frac{1}{\rho^2}\frac{\partial \rho}{\partial x_j}\frac{\partial p}{\partial x_k}\right)'' + \\ \omega_i''\epsilon_{ijk}\left(\frac{\partial}{\partial x_j}\left(\frac{1}{\rho}\frac{\partial \tau_{km}}{\partial x_m}\right)\right)'' \quad (C\,11)$$

Multiplying equation (C 11) by $\rho$, Reynolds-averaging, and then dividing by $\bar{\rho}/2$ gives

$$\frac{\partial \widetilde{\omega_i''\omega_i''}}{\partial t} - 2\omega_i''\frac{1}{\bar{\rho}}\frac{\partial \overline{\rho\widetilde{\omega_i u_j}}}{\partial x_j} + 2\omega_i''\frac{1}{\bar{\rho}}\frac{\partial \overline{\rho\widetilde{\omega_i}\widetilde{u_j}}}{\partial x_j} - 2\widetilde{\omega_i''\widetilde{u_j}}\frac{\partial \widetilde{\omega_i}}{\partial x_j} + 2\widetilde{\omega_i'' u_j\frac{\partial \omega_i}{\partial x_j}} = \\ 2\widetilde{\omega_i''\left(\omega_j\frac{\partial u_i}{\partial x_j}\right)''} - 2\widetilde{\omega_i''\left(\omega_i\frac{\partial u_j}{\partial x_j}\right)''} + 2\widetilde{\omega_i''\epsilon_{ijk}\left(\frac{1}{\rho^2}\frac{\partial \rho}{\partial x_j}\frac{\partial p}{\partial x_k}\right)''} + \\ 2\widetilde{\omega_i''\epsilon_{ijk}\left(\frac{\partial}{\partial x_j}\left(\frac{1}{\rho}\frac{\partial \tau_{km}}{\partial x_m}\right)\right)''} \quad (C\,12)$$

## REFERENCES

ABGRALL, R. & SAUREL, R. 2003 Discrete equations for physical and numerical compressible multiphase mixtures. *J. Comput. Phys.* **186** (2), 361–396.

BANUTI, D.T. 2015 Crossing the Widom-line – Supercritical pseudo-boiling. *The Journal of Supercritical Fluids* **98**, 12–16.

CASIANO, M. J., HULKA, J. R. & YANG, V. 2010 Liquid-propellant rocket engine throttling: A comprehensive review. *J. Propul. Power* **26** (5), 897–923.

CHUNG, T.-H., AJLAN, M., LEE, L. L. & STARLING, K. E. 1988 Generalized multiparameter correlation for nonpolar and polar fluid transport properties. *Ind. Engng. Chem. Fundam.* **27** (4), 671–679.

VAN DRIEST, E. R. 1951 Turbulent boundary layer in compressible fluids. *J. Aero. Sci.* **18**, 145–160, 216.

VAN DRIEST, E. R. 1956 On turbulent flow near a wall. *J. Aero. Sci.* **23**, 1007–1011, 1036.

FISHER, M. E. & WIDOM, B. 1969 Decay of correlations in linear systems. *J. Chem. Phys.* **50** (20), 3756–3772.

GORELLI, F., SANTORO, M., SCOPIGNO, T., KRISCH, M. & RUOCCO, G. 2006 Liquid-like behavior of supercritical fluids. *Phys. Rev. Lett* **97** (24), 245702.

HUANG, P. G., COLEMAN, G. N. & BRADSHAW, P. 1995 Compressible turbulent channel flows: DNS results and modelling. *J. Fluid Mech.* **305**, 185–218.

KAWAI, S. 2016 Direct numerical simulation of transcritical turbulent boundary layers at supercritical pressures with strong real fluid effects. In *54th AIAA Aerospace Sciences Meeting, San Diego, CA*, p. 1992.

KIM, KUKJIN, HICKEY, JEAN-PIERRE & SCALO, CARLO 2017*a* Numerical investigation of transcritical-t heat-and-mass-transfer dynamics in compressible turbulent channel flow. In *55th AIAA Aerospace Sciences Meeting*.

KIM, KUKJIN, SCALO, CARLO & HICKEY, JEAN-PIERRE 2017*b* Turbulent dynamics and heat transfer in transcritical channel flow. In *10th International Symposium on Turbulence and Shear Flow Phenomena*.

LARSSON, J., BERMEJO-MORENO, I. & LELE, S. K. 2013 Reynolds- and Mach-number effects in canonical shock–turbulence interaction. *J. Fluid Mech.* **717**, 293–321.

LARSSON, J. & LELE, S. K. 2009 Direct numerical simulation of canonical shock/turbulence interaction. *Phys. Fluids* **21**, 126101.

LARSSON, J., LELE, S. K. & MOIN, P. 2007 Effect of numerical dissipation on the predicted spectra for compressible turbulence. *Annual Research Briefs, Center for Turbulence Research, Stanford* .

LEMMON, E.W., MCLINDEN, M.O. & FRIEND, D.G. 2016 *Thermophysical Properties of Fluid Systems, NIST Chemistry WebBook, NIST Standard Reference Database*. Gaithersburg MD, 20899: National Institute of Standards and Technology.

MA, P. C., LU, Y. & IHME, M. 2017 An entropy-stable hybrid scheme for simulations of transcritical real-fluid flows. *J. Comp. Phys.* **340**, 330–357.

NEMATI, H., PATEL, A., BOERSMA, B. J. & PECNIK, R. 2015 Mean statistics of a heated turbulent pipe flow at supercritical pressure. *Intl. J. Heat Mass Transfer* **83**, 741–752.

PALUMBO, M. 2009 Predicting the Onset of Thermoacoustic Oscillations in Supercritical Fluids. PhD thesis, Purdue University.

PATEL, A., PEETERS, J. W. R., BOERSMA, B. J. & PECNIK, R. 2015 Semi-local scaling and turbulence modulation in variable property turbulent channel flows. *Phys. Fluids* **27**, 095101.

PEETERS, J. W. R., PECNIK, R., ROHDE, M., VAN DER HAGEN, T. H. J. J. & BOERSMA, B. J. 2016 Turbulence attenuation in simultaneously heated and cooled annular flows at supercritical pressure. *J. Fluid Mech.* **799**, 505–540.

PENG, D.-Y. & ROBINSON, D. B. 1976 A new two-constant equation of state. *Ind. Engng. Chem. Fundam.* **15** (1), 59–64.

PIZZARELLI, M., NASUTI, F., PACIORRI, R. & ONOFRI, M. 2009 Numerical analysis of three-dimensional flow of supercritical fluid in asymmetrically heated channels. *AIAA Journal* **47**, 2534–2543.

POLING, B. E., PRAUSNITZ, J. M. & O'CONNELL, J. P. 2001 *The Properties of Gases and Liquids*. McGraw-Hill.

Sengers, J. V., Kayser, R. F., Peters, C. J., H. J. White, Jr., Ewing, M. B., Trusler, J. P. M., Anderko, A., Boublik, T., Kalyuzhnyi, Yu. V., Cummings, P. T., Smirnova, N. A., Victorov, A. V., Ely, J. F., Marrucho, I. M. F., Sandler, S. I., Orbey, H., Matteoli, E., Hamad, E. Z., Mansoori, G. A. & Anisimov, M. A. 2000 *Equation of state for fluids and fluid mixtures. Part I*, 1st edn. Elsevier.

Sengupta, U., Nemati, H., Boersma, B. J. & Pecnik, R. 2017 Fully compressible low-mach number simulations of carbon-dioxide at supercritical pressures and trans-critical temperatures. *Flow Turbulence Combust* **99**, 909–931.

Simeoni, G. G., Bryk, T., Gorelli, F. A., Krisch, M., Ruocco, G., Santoro, M. & Scopigno, T. 2010 The Widom line as the crossover between liquid-like and gas-like behaviour in supercritical fluids. *Nature Physics* **6**, 503–507.

Terashima, H., Kawai, S. & Yamanishi, N. 2011 High-resolution numerical method for supercritical flows with large density variations. *AIAA Journal* **49**, 2658–2672.

Terashima, H. & Koshi, M. 2012 Approach for simulating gas/liquid-like flows under supercritical pressures using a high-order central differencing scheme. *J. Comput. Phys.* **231** (20), 6907–6923.

Terashima, H. & Koshi, M. 2013 Strategy for simulating supercritical cryogenic jets using high-order schemes. *Comput. Fluids* **85**, 39–46.

The HDF Group 1998 Hierarchical Data Format, version 5. Http://www.hdfgroup.org/HDF5/.

Thurston, R. S. 1964 Pressure Oscillations Induced by Forced Convection Heat Transfer to Two Phase and Supercritical Hydrogen. *Tech. Rep.* LAMS-3070. Los Alamos Scientific Laboratory.

Trettel, A. & Larsson, J. 2016 Mean velocity scaling for compressible wall turbulence with heat transfer. *Phys. Fluids* **28**, 026102.

Tucker, S. C. 1999 Solvent density inhomogeneities in supercritical fluids. *Chem. Rev.* **99**, 391–418.

Wang, H., Zhou, J., Pan, Y. & Wang, N. 2015 Experimental investigation on the onset of thermo-acoustic instability of supercritical hydrocarbon fuel flowing in a small-scale channel. *Acta Astronautica* **117**, 296–304.

Wen, Q. L. & Gu, H. Y. 2011 Numerical investigation of acceleration effect on heat transfer deterioration phenomenon in supercritical water. *Progr. Nucl. Energy* **53**, 480–486.

Yoo, J. Y. 2013 The turbulent flows of supercritical fluids with heat transfer. *Annu. Rev. Fluid Mech.* **45**, 495 – 525.

Zhang, L., Liu, M., Dong, Q. & Zhao, S. 2011 Numerical research of heat transfer of supercritical $CO_2$ in channels. *Energy Power Engng.* **3**, 167–173.

Zhong, F., Fan, X., Yu, G., Li, J. & Sung, C.-J. 2009 Heat transfer of aviation kerosene at supercritical conditions. *J. Thermophys. Heat Transfer* **23**, 543–550.